\documentclass[sigconf, nonacm]{acmart}

\AtBeginDocument{%
  }

\usepackage{xcolor}
\usepackage{enumitem}
\usepackage{tcolorbox}

\newcommand{\toolname}{\textit{AniMaster}}
\newcommand{\story}{\textit{Story Space}}
\newcommand{\script}{\textit{Script Space}}
\newcommand{\Video}{\textit{Video Space}}

\definecolor{envbg}{RGB}{61, 110, 92}
\definecolor{actbg}{RGB}{140, 70, 84}
\definecolor{reabg}{RGB}{74, 100, 136}
\definecolor{diabg}{RGB}{122, 104, 56}
\definecolor{revbg}{RGB}{106, 82, 128}
\definecolor{paubg}{RGB}{82, 90, 102}
\definecolor{prebg}{RGB}{26, 138, 154}

\newcommand{\BeatENV}{\tcbox[on line, colback=envbg!10, colframe=envbg!10, boxrule=0pt, arc=3pt, boxsep=0pt, left=2pt, right=2pt, top=1pt, bottom=1pt]{\textcolor{envbg}{Environment}}}
\newcommand{\BeatACT}{\tcbox[on line, colback=actbg!10, colframe=actbg!10, boxrule=0pt, arc=3pt, boxsep=0pt, left=2pt, right=2pt, top=1pt, bottom=1pt]{\textcolor{actbg}{Action}}}
\newcommand{\BeatREA}{\tcbox[on line, colback=reabg!10, colframe=reabg!10, boxrule=0pt, arc=3pt, boxsep=0pt, left=2pt, right=2pt, top=1pt, bottom=1pt]{\textcolor{reabg}{Reaction}}}
\newcommand{\BeatDIA}{\tcbox[on line, colback=diabg!10, colframe=diabg!10, boxrule=0pt, arc=3pt, boxsep=0pt, left=2pt, right=2pt, top=1pt, bottom=1pt]{\textcolor{diabg}{Dialogue}}}
\newcommand{\BeatREV}{\tcbox[on line, colback=revbg!10, colframe=revbg!10, boxrule=0pt, arc=3pt, boxsep=0pt, left=2pt, right=2pt, top=1pt, bottom=1pt]{\textcolor{revbg}{Reveal}}}
\newcommand{\BeatPAU}{\tcbox[on line, colback=paubg!10, colframe=paubg!10, boxrule=0pt, arc=3pt, boxsep=0pt, left=2pt, right=2pt, top=1pt, bottom=1pt]{\textcolor{paubg}{Pause}}}
\newcommand{\BeatPRE}{\tcbox[on line, colback=prebg!10, colframe=prebg!10, boxrule=0pt, arc=3pt, boxsep=0pt, left=2pt, right=2pt, top=1pt, bottom=1pt]{\textcolor{prebg}{Presentation}}}

            \newcommand{\parti}[1]{\textbf{#1}}

\DeclareRobustCommand{\iconref}[1]{\raisebox{-0.2\height}{\includegraphics[height=0.9em]{icons/#1.png}}}
\newcommand{\uibox}[1]{\tcbox[on line, colback=gray!12, colframe=gray!12, boxrule=0pt, arc=3pt, boxsep=0pt, left=2pt, right=2pt, top=1pt, bottom=1pt]{#1}}
\newcommand{\maincanvas}{\uibox{\textit{Main Canvas}}}
\newcommand{\scriptreader}{\uibox{\textit{Script Reader}}}
\newcommand{\assetlib}{\uibox{\textit{Asset Library}}}

\newcommand{\inspector}{\uibox{\textit{Inspector}}}
\newcommand{\canvasmap}{\uibox{\textit{Canvas Map}}}
\newcommand{\overview}{\uibox{\textit{Overview}}}
\newcommand{\breakdown}{\uibox{\textit{Breakdown}}}
\newcommand{\beatcanvas}{\uibox{\textit{Beat Canvas}}}
\newcommand{\frameinspector}{\uibox{\textit{Frame Inspector}}}
\newcommand{\videoinspector}{\uibox{\textit{Video Inspector}}}
\newcommand{\videoeditor}{\uibox{\textit{Video Editor}}}
\newcommand{\compositioneditor}{\uibox{\textit{Composition Editor}}}
\newcommand{\variantsview}{\uibox{\textit{Variants View}}}
\newcommand{\beatsuggestion}{\uibox{\textit{Beat Suggestion}}}
\newcommand{\cinematiccue}{\uibox{\textit{Cinematic Cue}}}

\begin{document}

\title{AniMaster: From Story Texts to Animated Videos via Cinematic Script Generation and Interactive Authoring}

\author{Ruiqi Yu}
\authornote{These authors contributed equally to this work.}
\email{ruiqiyu@hdu.edu.cn}
\orcid{0000-0001-6116-6678}
\affiliation{%
  \institution{Hangzhou Dianzi University}
  \city{Hangzhou}
  \state{Zhejiang}
  \country{China}
}
\affiliation{%
  \institution{Nanyang Technological University}
  \city{Singapore}
  \country{Singapore}
}

\author{Dekun Qian}
\authornotemark[1]
\email{qiandekun@hdu.edu.cn}
\affiliation{%
  \institution{Hangzhou Dianzi University}
  \city{Hangzhou}
  \state{Zhejiang}
  \country{China}
}

\author{Jiale Xu}
\email{xujiale@hdu.edu.cn}
\affiliation{%
  \institution{Hangzhou Dianzi University}
  \city{Hangzhou}
  \state{Zhejiang}
  \country{China}
}

\author{Sizhe Cheng}
\email{sizhe003@e.ntu.edu.sg}
\affiliation{%
  \institution{Nanyang Technological University}
  \city{Singapore}
  \country{Singapore}
}

\author{Yize Li}
\email{liyize1996@126.com}
\affiliation{%
  \institution{Hangzhou Dianzi University}
  \city{Hangzhou}
  \state{Zhejiang}
  \country{China}
}

\author{Xiangyang Wu}
\email{wuxy@hdu.edu.cn}
\affiliation{%
  \institution{Hangzhou Dianzi University}
  \city{Hangzhou}
  \state{Zhejiang}
  \country{China}
}

\author{Zhiguang Zhou}
\email{zhgzhou@hdu.edu.cn}
\affiliation{%
  \institution{Hangzhou Dianzi University}
  \city{Hangzhou}
  \state{Zhejiang}
  \country{China}
}

\author{Wei Chen}
\email{chenvis@zju.edu.cn}
\affiliation{%
  \institution{Zhejiang University}
  \city{Hangzhou}
  \state{Zhejiang}
  \country{China}
}

\author{Yong Wang}
\email{yong-wang@ntu.edu.sg}
\affiliation{%
  \institution{Nanyang Technological University}
  \city{Singapore}
  \country{Singapore}
}

\renewcommand{\shortauthors}{Yu et al.}

\begin{abstract}
Recent advances in Video Generation Models (VGMs) have demonstrated strong capabilities in producing short video clips.
However, it is still challenging for everyday creators to leverage these models to produce polished long-form animated videos from brief story texts. Informed by a formative study with both novice creators and film experts, we identify two major challenges of interactive video authoring: (1) the lack of expertise in translating free-form story texts to professional cinematic scripts and finally high-quality animated videos, and (2) the absence of effective ways to convey video design intents to key variables of visual storytelling, such as shot composition, camera controls and shot sequencing.
Drawing on narratology and film studies, we propose a three-layer design framework that defines the key design dimensions across three layers (i.e., story texts, cinematic scripts, and animated videos) as well as the translation between them.
Built on this framework, we present \toolname{}, a VGM-powered authoring tool to enable everyday creators to easily produce smooth animated videos from free-form story texts.
\toolname{} automatically expands brief story texts to detailed cinematic scripts, and further translates cinematic scripts into polished videos by following professional visual storytelling principles. It also allows users to interactively edit the scripts and refine the generated videos via text instructions and intuitive interactions. 
We extensively evaluated \toolname{} through an in-depth user study with 16 participants, two case studies, and expert interviews with 2 film professionals. 
The results demonstrate the effectiveness and usability of \toolname{} in helping everyday creators create polished animated videos from free-form story texts. 

\end{abstract}


\begin{teaserfigure}
  \includegraphics[width=\textwidth]{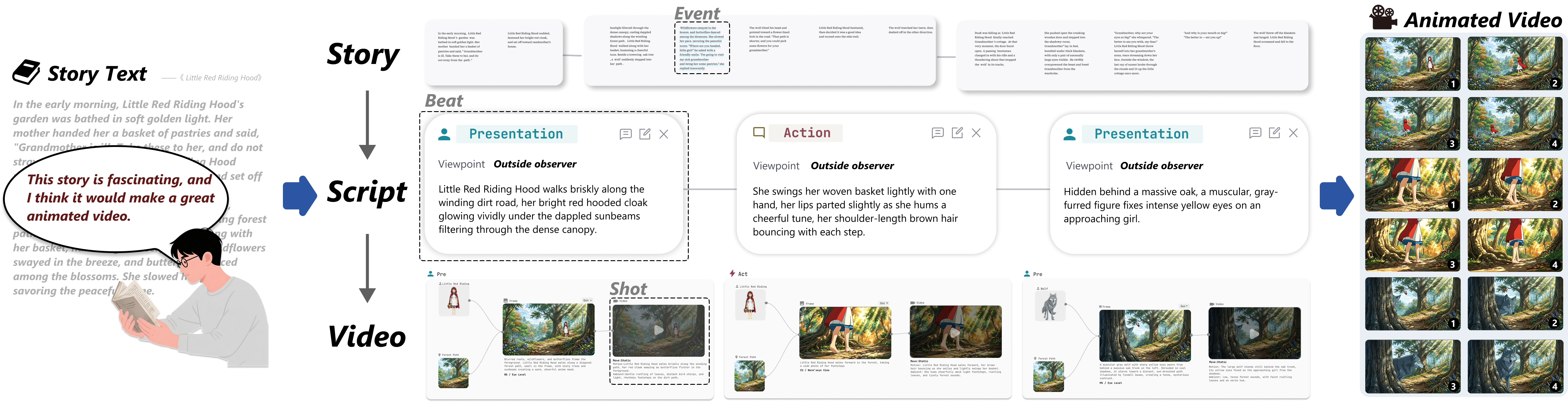}
  \caption{
    \toolname{} enables everyday creators to produce animated videos from free-form story texts.
        A creator inputs a \textit{Story Text} (left), which is automatically structured as a sequence of \textit{Events}.
        Each \textit{Event} is then translated into \textit{Beats} annotated with cinematic attributes, bridging narrative structure and film language (middle).
        The creator composes \textit{Shots} for each \textit{Beat} and generates the final \textit{Animated Video} (right).
      }
  \label{fig:teaser}
\end{teaserfigure}

\maketitle

\section{Introduction}

Video generation models (VGMs) have shown impressive capabilities in generating short video clips in the past few years~\cite{ho2022video, blattmann2023stable, wan2025wan}. 
With the explosive popularity of video-based social media platforms such as TikTok\footnote{\url{https://www.tiktok.com/}}, YouTube\footnote{\url{https://www.youtube.com/}} and Bilibili\footnote{\url{https://www.bilibili.com/}}, there is a fast-growing interest and need from common social media users to efficiently and easily create high-quality animated videos.
Despite the recent advancement of VGMs, it is still highly challenging for everyday creators, who do not receive professional training in filmmaking, to utilize existing video models and easily produce polished long-form animated videos from brief story texts~\cite{yu2024barriers,liu2026text,elmoghany2025survey}.

Recently, there has been research effort in leveraging Artificial Intelligence (AI) techniques to facilitate video generation. 
For example, Doki~\cite{liu2026text} integrates AI agents into a text-native interface to help everyday users to achieve video authoring via text descriptions, and Vidmento~\cite{yeh2026vidmento} allows users to combine captured and AI-generated media to interactively create hybrid videos.
However, these approaches do not reflect the professional workflow of videographer or filmmakers, and have not incorporated the foundational narratology and cinematic guidelines, making them unable to enable everyday users to easily produce polished long-form videos.
AI researchers have also applied generative AI models to support video style transfer~\cite{wei2025cinevision, huang2025filmaster}, which intrinsically requires existing reference videos and results in barriers for them to be adopted by everyday users. 
The latest VGMs have also started to support sophisticated camera parameter configurations in video generation~\cite{google2025veo3, wan2025wan, bytedance2026seedance2}, which is, however, usually not accessible enough for everyday creators without professional expertise and hard for them to translate their creative intents into precise prompt instructions.
Despite all these efforts, an effective approach to enable everyday users to easily produce polished long-form videos from free-form story texts is still urgently needed.

To gain a deeper understanding of the research gap, we conducted a formative study with both 12 novice creators and 6 film experts. We found that creating videos from story texts is not an end-to-end process, but requires a detailed workflow to translate free-form story texts to professional scripts and further to final desired videos, which needs users to consider crucial factors like shot composition and camera controls.
However, such a workflow poses two significant challenges to everyday users:
\textbf{(P17)} the translation among free-form story texts, scripts and final animated videos relies on professional conventions and expertise, and everyday creators lack such expertise to achieve effective translations; 
\textbf{(P18)} everyday creators have creative intents but there are no convenient ways for them to convert these intents into video parameters such as shot composition, camera controls, and shot sequencing.

To address these challenges, drawing on narratology~\cite{bal2017narratology, mckee1997story, genette1980narrative} and film studies ~\cite{bordwell2019film, katz1991film}, we propose a three-layer design framework that defines the key design dimensions across three layers—story texts, cinematic scripts, and animated videos—as well as the translation principles between them.
Built on this framework, we develop \toolname{}\footnote{\url{https://animaster-tool.github.io/}}, a novel authoring tool that integrates Large Language Models (LLMs) and VGM to enable everyday creators to easily produce smooth animated videos from free-form story texts. \toolname{} operationalizes the framework through two stages of translation rules: it first expands story events into editable beat sequences in the cinematic script layer, and then translates beats into shots with executable visual parameters in the video layer. In this way, \toolname{} exposes to users the intermediate narrative and cinematic decisions that are usually implicit in professional practices \textbf{(P17)}.
Also, 
\toolname{} provides a canvas-based interface to show automatically generated global narrative structure, beat-level planning, and shot-level video preview.
At each level, users can inspect and revise the system's outputs with intuitive interactions: they can modify the generated beat structure, adjust shot parameters such as composition and camera movement, and express higher-level revision intents through natural language or direct manipulation. 
By combining rule-guided generation with interactive editing, \toolname{} helps users externalize narrative intent and iteratively refine how a story should be conveyed visually \textbf{(P18)}.

{We extensively evaluated \toolname{} through an in-depth user study with 16 participants, two case studies, and expert interviews with 2 film professionals. The results demonstrate the effectiveness and usability of \toolname{} in helping everyday creators create polished animated videos from free-form story texts. More broadly, they suggest the value of making professional narrative and cinematic knowledge explicit, editable, and computationally actionable in AI-supported creative tools.}

In summary, our contributions are as follows:
\begin{itemize}[leftmargin=*]
    \item \textbf{A three-layer design framework.} Grounded in narratology and film studies, we systematically distill the key design dimensions across story texts, cinematic scripts, and animated videos, as well as the translation rules between them.
    \item \textbf{The \toolname{} authoring tool.} \toolname{} operationalizes the framework through rule-guided multi-stage translation and interactive editing, enabling everyday creators to progressively transform free-form story texts into animated videos.
    \item \textbf{An extensive evaluation.} Through a user study, case studies, and expert interviews, we show how structured intermediate representations and translation rules can lower the barrier of cinematic video creation for non-expert creators.
\end{itemize}

\section{Related Work}

We organize related work along the story-to-cinematic-video pipeline into 3 groups:
(1)~narrative and scriptwriting, which asks how to make textual narratives more structured and controllable;
(2)~pre-visualization and authoring interfaces, which asks how to let creators iterate rapidly at the shot level; and
(3)~generative cinematic production, which asks how to push high-level creative intent toward producible video.

\subsection{Narrative and Scriptwriting}

In the cross-modal direction, researchers have introduced visual retrieval, narrative-structure visualization, and automatic script parsing to support preliminary text-to-visual mapping~\cite{rao2024scriptviz,masson2025visual,kato2024griffith,jorgensen2023screenplay,tian2025large}.
Recently, Doki~\cite{liu2026text} maps documents, paragraphs, and sentences to videos, sequences, and shots, enabling text to serve simultaneously as a narrative vehicle and an executable script.
Although works such as Doki have begun to bridge text and video generation directly, existing approaches still lack a systematic encoding of narratological and cinematic knowledge into an intermediate representation accessible to creators; the semantic gap from script to executable visuals remains unresolved.

\subsection{Pre-visualization and Authoring Interfaces }

At the visual-representation level, the core challenge is maintaining character and environment consistency across storyboard frames.
Recent work has made notable progress through identity refinement, consistency attention, and pose-guided generation~\cite{avrahami2024chosen,zhou2024storydiffusion,hu2024animate}.
Moving to spatial and temporal planning, 3D engines have extended pre-visualization to automatic camera control, text-driven 3D previsualization, and composition optimization~\cite{he2023virtual,kim2021asap,chen2024cinepregen,he2024interactive}; underpinning such systems, Jhala et al.~\cite{jhala2005discourse} formalized \emph{film idioms} that map narrative functions to camera expressions.
On the interaction paradigm side, canvas-based interfaces are increasingly adopted for creative workflows, enabling creators to organize materials and ideas spatially rather than sequentially~\cite{chung2024patchview,suh2024luminate,brade2023promptify}, and some tools further combine freeform canvases with sequential UIs such as timelines or text editors~\cite{zhou2024visar,cao2025compositional}.
However, existing pre-visualization tools, while advancing individual components, lack explicit modeling of long-range narrative logic. Canvas-based tools likewise focus on freeform ideation and material organization, without exploring how spatial layout can guide narrative flow. Creators still cannot simultaneously oversee narrative structure and orchestrate shot language within a unified framework.

\subsection{Generative Cinematic Production}

The diffusion paradigm has provided a foundation for video generation. Research has been advancing along 2 axes, generation quality and temporal coherence~\cite{ho2022video,blattmann2023stable,wan2025wan,google2025veo3}, and recent industrial models further support cross-scene identity and style consistency~\cite{bytedance2026seedance2}.
Another line of work chains script planning, shot design, and keyframe generation into end-to-end cinematic pipelines~\cite{li2024anim,zhang2025bridging,huang2025filmaster};
Vidmento~\cite{yeh2026vidmento} applies cinematography-informed generative expansion to existing footage organized at the story, scene, and shot levels, giving creators a structured canvas for cinematic assembly.
To make cinematic language computable, MovieNet~\cite{huang2020movienet} and Stable Cinemetrics~\cite{chatterjee2025stable} have respectively built a large-scale film-understanding dataset and a control taxonomy, while work on shot-scale classification, shot-boundary detection, and camera-trajectory editing continues to supply control entry points for production contexts~\cite{savardi2018shot,soucek2024TransNetv2,he2025cameractrlenablingcameracontrol}.
Yet in practice these capabilities are still used as black boxes: creators cannot stably constrain shot scheduling and montage logic with directorial thinking, and narrative coherence and pacing at the sequence level remain difficult to guarantee.
\begin{figure*}[t]
  \centering
  \includegraphics[width=\textwidth]{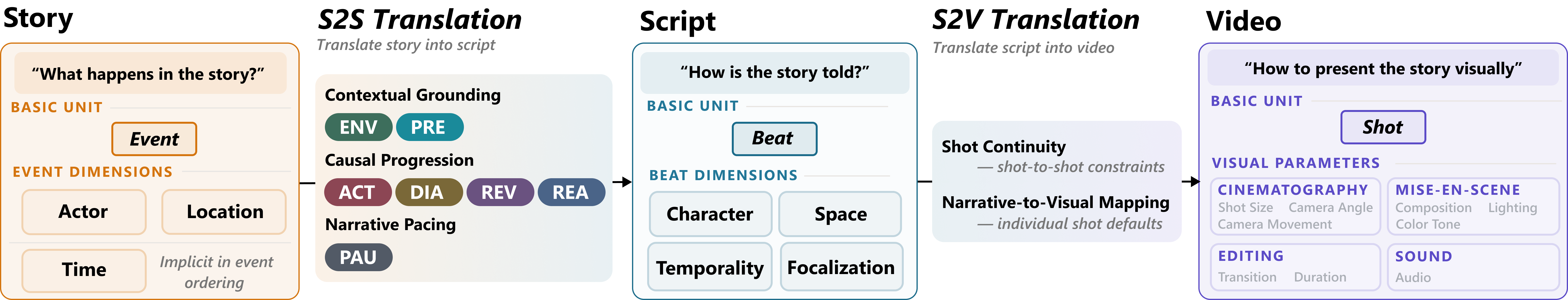}
  \caption{The three-layer design framework. \textbf{Story Space} defines Events; \textbf{Script Space} defines Beats with 4 narrative dimensions and seven Beat Types; \textbf{Video Space} defines Shots with nine visual parameters. Translation Logic~1 maps Events to Beats; S2V Translation maps Beats to Shots.}
  \label{fig:pipeline}
  \label{fig:framework}
\end{figure*}

\section{Design Framework}

We begin with a formative study to identify the core design challenges, and then derive a three-layer design framework grounded in narratology and film studies.

\subsection{Formative Study}

We conducted a two-phase formative study to understand the key difficulties non-professional creators face when translating stories into visual sequences. The two phases were user observation and expert interview.

\subsubsection{Study Procedure}\mbox{}

\textbf{User observation.}
We recruited 12 participants with no filmmaking background and asked them to transform a roughly 300-word story text into a sequence of visual frames using existing AI video generation tools, while following a think-aloud protocol. Our observation focused on recording the specific failure patterns and cognitive bottlenecks that participants encountered during the transformation process.

 \textbf{Expert interview.}
To explain the failure patterns identified during user observation from a professional perspective, we invited 6 experts with extensive filmmaking experience (\parti{E1}--\parti{E6}) for semi-structured interviews of approximately 60 minutes each. Among them, \parti{E1}--\parti{E2} have won awards at international AI film festivals, \parti{E3} is a content creator who teaches directing knowledge online, and \parti{E4}--\parti{E6} are AI-assisted creation bloggers with over 10,000 followers each. During the interviews, we presented the typical failure patterns recorded during the user observation phase and asked the experts to analyze the root causes of these failures, as well as to describe their thought processes when performing the same task.

We coded and categorized the behavioral data from user observations, and conducted a thematic analysis~\cite{braun2006using} on the expert interview data. This process yielded three design challenges. 
\subsubsection{Professional Pipeline of Animated Video Generation}

Our analysis revealed that professional animated video creation does not proceed directly from story text to final visuals. 

In the user observation, many participants tried to move directly from free-form story text to visual frames. They could often generate individual images or short clips, but had trouble deciding what to show first, how to break a story event into smaller moments, and how adjacent shots should connect. As a result, their outputs were often locally reasonable but lacked clear progression and coherence.

In the expert interviews, we presented these failure patterns and asked experts to explain their causes and describe how they themselves would approach the same task. Across interviews, a common workflow emerged. Experts first identify what happens in the story, then decide how the story should be told to the audience, and finally turn those decisions into concrete shots. As \textbf{\parti{E1}} explained, \textit{``After I get a script, I first figure out what happens, then think about how to tell it, and finally decide how to shoot it. The thinking is completely different at each step.''}

Taken together, these findings suggest a three-step professional pipeline for animated video generation, moving from \textbf{Story} to \textbf{Script} and then to \textbf{Video}. This pipeline helps explain why non-expert creators struggle with current tools. We next distill the main design challenges revealed by this pipeline.

\subsubsection{Design Challenges}

Based on the failure patterns from user observation and the professional analysis from expert interviews, we identified two major design challenges for everyday users:

\textbf{DC1: The lack of expertise in translating free-form story texts to professional cinematic scripts and videos.} 
During user observation, most novice participants systematically failed when they attempted to go directly from story text to visual frames. Experts revealed the reason: transforming a story into visuals is not a single step, but requires multiple stages of design, such as understanding the story, orchestrating the narrative, and designing shots---each demanding a fundamentally different mode of thinking.
The transitions between stages follow professional conventions that non-experts cannot access. Participants' outputs lacked shot sequencing logic; experts noted these conventions require long-term accumulation. As \parti{E3} described: \textit{``Many design choices of his professional films come from years of filmmaking and film-watching experiences.''}

\textbf{DC2: The absence of effective ways to convey
video design intents to key variables of visual storytelling.}
Participants without video creation expertise could articulate desired narrative effects (e.g. ``tension'') but could not map them to visual parameters such as shot composition or camera controls. As \parti{P2} noted:\textit{``I want the audience to feel the character's tension, but I don't know what kind of shot to use to convey that feeling.''}

\subsection{Three-Layer Structure}
Our formative study shows that directors follow a multi-stage cognitive process. To identify and formalize this process, we surveyed the narratology and film studies literature.
 
In narratology, Genette~\cite{genette1980narrative}, Chatman~\cite{chatman1978story}, and Bal~\cite{bal2017narratology} each use different terms but point to a shared insight: between a story's raw events and its final presentation lies an independent orchestration layer, where creators decide in what order and from what perspective the audience receives information. In film studies, Bordwell~\cite{bordwell1985narration} argues that stylistic choices serve narrative functions, and Katz~\cite{katz1991film} details how narrative functions map to camera setups. Together, both disciplines reach the same conclusion: going from story to visuals requires two transitions, with an intermediate layer that encodes narrative orchestration decisions in between.
 
Based on this, we organize the design framework into three Spaces. \textbf{\story{}} represents the logical content of story events. \textbf{\script{}} represents audience-oriented narrative orchestration strategies. \textbf{\Video{}} represents executable visual parameters.

\subsection{Space Dimensions}

The three-layer structure establishes the skeleton of the design framework. We derive dimensions in two directions: top-down, by identifying candidate concepts from narratology and film studies that each layer should carry; and bottom-up, by tracing what information each layer must deliver to downstream VGMs. Because not every theoretical concept is suitable for operationalization, we apply three screening criteria throughout the derivation: \textbf{Extractability}, whether the dimension can be automatically extracted from text; \textbf{Operability}, whether everyday creators can understand and edit it; and \textbf{Renderability}, whether it can be compiled into parameters that VGMs can execute.

\subsubsection{Story Space}

\story{} represents \emph{what happens in the story}. We draw on Bal's concept of \emph{fabula} to build this layer~\cite{bal2017narratology, masson2025visual}. Bal defines a fabula as a logical layer with four basic elements: Actor, Location, Time, and Event~\cite{bal2017narratology}. Event is the most self-contained among them. An Event naturally includes  \textbf{Actor} and  \textbf{Location} as attributes, and  \textbf{Time} is expressed implicitly through the ordering of Events. This makes Event the natural basic unit of \story{}~\cite{liveley2019narratology}.

\subsubsection{Script Space}

\script{} represents \emph{how the story is told}. It reorganizes story facts into narrative decisions oriented toward the audience's experience. The basic unit is the \textbf{Beat}, the smallest element of structure that carries a single narrative function~\cite{mckee1997story}. One Event typically unfolds into an ordered sequence of several Beats. Each Beat also has a \textbf{Beat Type} label that identifies its narrative function (e.g. establishing the environment, depicting an action). 

Bal~\cite{bal2017narratology} organizes narrative orchestration along 4 aspects, which we adopt as the dimensions of each Beat: \textbf{Temporality} (covering temporal ordering, pacing, and frequency), \textbf{Characters}, \textbf{Space} (referring to the narrative space dimension, not the layer naming in our framework), and \textbf{Focalization} (the perspective from which events are perceived, further divided into zero, internal, and external focalization following Genette~\cite{genette1980narrative}).

\subsubsection{Video Space}

\Video{} represents \emph{how to present the story visually}. It turns narrative decisions into concrete visuals that the audience can see. Bordwell divides the stylistic system into four subsystems: Mise-en-sc\`ene, Cinematography, Editing, and Sound~\cite{bordwell1985narration}. Katz further defines the \emph{camera setup} as the basic unit of visual execution~\cite{katz1991film}. Accordingly, we adopt the \textbf{Shot} as the basic unit of \Video{}, and organize the key variables of visual storytelling into 9 parameters across the four subsystems:

\textbf{Cinematography} covers Shot Size, Camera Angle, and Camera Movement. \textbf{Mise-en-sc\`ene} covers Composition, Lighting, and Color Tone. \textbf{Editing} covers Transition and Duration. \textbf{Sound} covers Audio.

\begin{figure*}[t]
  \centering
  \includegraphics[width=\textwidth]{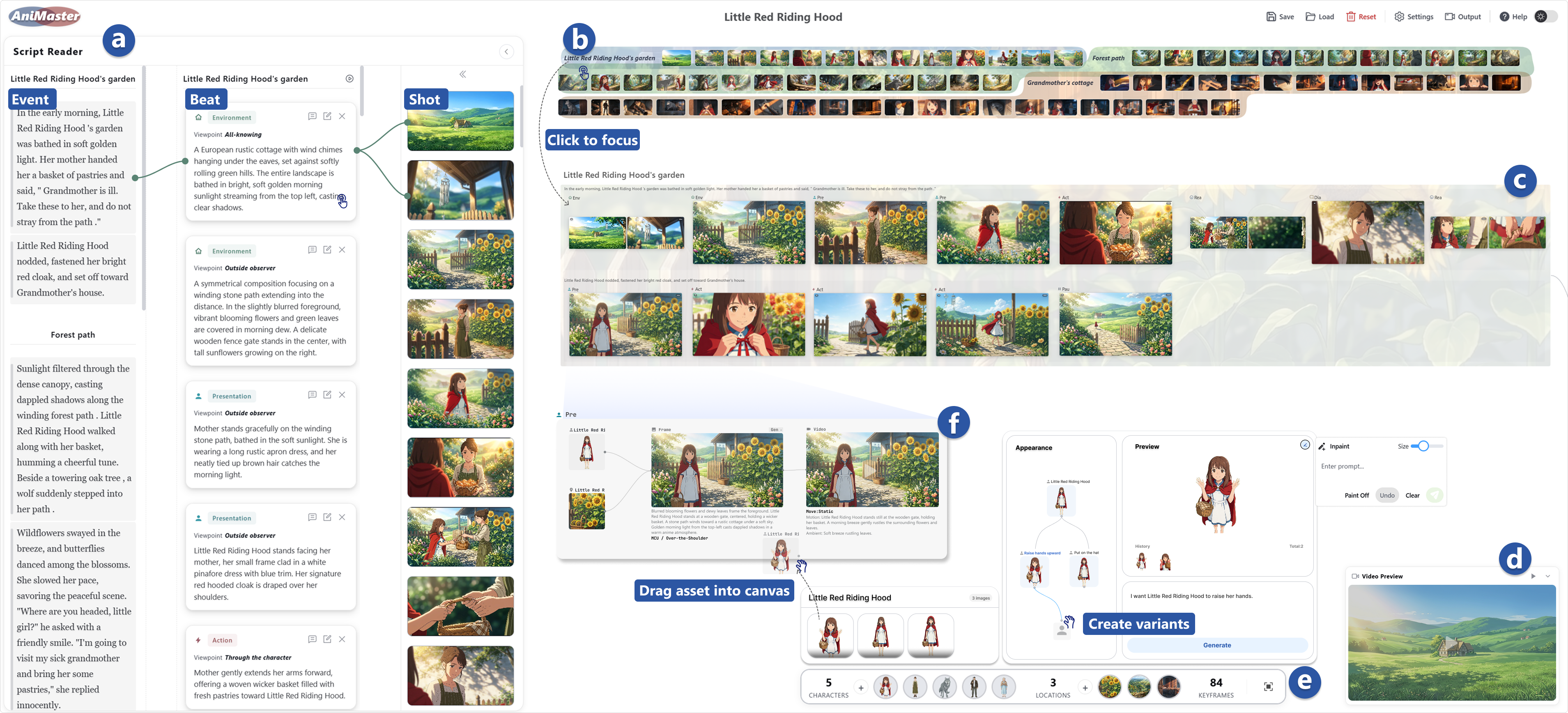}
  \caption{
    \toolname{} is an LLM-powered authoring interface for progressively creating cinematic animations from story texts. \iconref{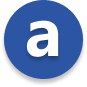}~The \scriptreader{} on the left aligns story, script, and video text across layers. The \maincanvas{} on the right hosts the primary authoring workspace, where creators \iconref{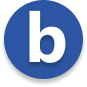}~navigate shots globally via the \canvasmap{}, \iconref{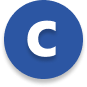}~review Beat sequences in the \breakdown{}, \iconref{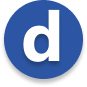}~preview the assembled animation, \iconref{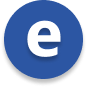}~manage reusable assets in the \assetlib{}, and \iconref{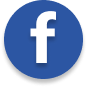}~compose individual shots in the \beatcanvas{}.
  }
  \label{fig:system-overview}
\end{figure*}

\subsection{Translation}

DC2 shows that transitions between stages rely on professional conventions that remain implicit. The translation makes these conventions explicit and computable. 

\subsubsection{From Story to Script}

The core task in going from \story{} to \script{} is to unfold each Event into an ordered sequence of Beats. Following Bordwell~\cite{bordwell1985narration}, who argues that viewers actively construct meaning along space, time, and causality, we define three \textbf{Narrative Unfolding Conditions}:

\begin{enumerate}[label=\textbf{(\arabic*)}, leftmargin=*]
    \item \textbf{Contextual Grounding.} In classical narrative cinema, a new scene typically begins by showing the space before introducing characters, giving the audience basic context~\cite{bordwell2006way}. When the scene or characters change, the context needs to be set up again.

    \item \textbf{Causal Progression.} Bordwell sees causality as the main force driving how audiences make sense of a story~\cite{bordwell1985narration}. The narrative moves forward through actions and dialogue, and each step needs a causal link to the next. For example, an action calls for a reaction (the Kuleshov effect~\cite{kuleshov1974kuleshov}), and a reveal needs prior setup (Katz's question-and-answer pattern~\cite{katz1991film}).

    \item \textbf{Narrative Pacing.} After a series of narrative actions, the audience needs room to absorb what has happened~\cite{mckee1997story}. Pacing controls the viewing experience by alternating between forward movement and pauses.
\end{enumerate}

Accordingly, we define seven Beat Types. \BeatENV\linebreak
and \BeatPRE serve Contextual Grounding: the former establishes the spatial setting and the latter introduces characters. \BeatACT, \BeatDIA, and \BeatREV push the causal chain forward (corresponding to Barthes' cardinal functions~\cite{barthes1966introduction}), while \BeatREA closes the loop by responding to preceding events. \BeatPAU serves Narrative Pacing, giving the audience a moment to process.

\subsubsection{From Script to Video}

The core task in going from \script{} to \Video{} is to turn each Beat into executable visual parameters. Bordwell~\cite{bordwell1985narration} and Katz~\cite{katz1991film} show that stylistic choices serve narrative functions and that camera setups have systematic correspondences with narrative intent. Based on this, we define 2 \textbf{Visual Translation Principles}:
\begin{enumerate}[label=\textbf{(\arabic*)}, leftmargin=*]
    \item \textbf{Shot Continuity.} Viewers rely on a sense of spatial and temporal continuity across consecutive shots to follow the narrative. When this continuity breaks, they feel pulled out of the story~\cite{bordwell2006way}. Classical continuity editing~\cite{bordwell2006way,katz1991film} offers a set of well-tested conventions for this. On one hand, certain narrative content calls for certain shot patterns (e.g. dialogue uses shot-reverse-shot). On the other hand, adjacent shots must meet visual constraints (e.g. maintaining a consistent axis) to preserve the audience's sense of space.
    
    \item \textbf{Narrative Function Mapping.} Different narrative units serve different cognitive purposes, so they should look different on screen. For example, establishing a setting favors long and wide shots, while capturing emotional reactions favors close-ups. The Beat Type determines both the shot pattern and the number of shots, and provides default values for each shot's visual parameters.
\end{enumerate}

These two principles work together. Shot Continuity governs how shots relate to each other, while Narrative Function Mapping decides what each individual shot looks like.

\section{\toolname{}}
 
Built on the three-layer design framework, we designed and implemented \toolname{}, a system that helps everyday creators progressively build cinematic animations from story texts.

\subsection{Interface Overview}

The \toolname{} interface consists of the \scriptreader{} on the left and the \maincanvas{} on the right (\autoref{fig:system-overview}). The \scriptreader{} (\autoref{fig:system-overview}\iconref{a}) provides a cross-layer view in three aligned columns: \textit{Events in Story} on the left, \textit{Beats} in the middle, and \textit{Shots} on the right, letting creators trace how a story moment is expanded and realized across layers. The \maincanvas{} hosts the primary authoring workspace and integrates five components: the \canvasmap{} (\autoref{fig:system-overview}\iconref{b}) provides a global thumbnail overview for quick shot-level navigation; the \breakdown{} view (\autoref{fig:system-overview}\iconref{c}) reveals the Beat sequence within each Event through semantic zooming; the Video Preview panel (\autoref{fig:system-overview}\iconref{d}) lets creators review the assembled animation at any time; the \assetlib{} (\autoref{fig:system-overview}\iconref{e}) manages reusable character and location variants that can be dragged onto the canvas; and the \beatcanvas{} (\autoref{fig:system-overview}\iconref{f}) opens a dedicated free-form workspace for composing individual shots within a Beat. Through semantic zooming on the \maincanvas{}, creators can progressively move from global narrative structure (\overview{}) through Beat-level planning (\breakdown{}) to shot-level production (\beatcanvas{}), matching the three design layers.

\subsection{Feature Walkthrough}
 
Next, we follow a creator’s perspective as they turn \textit{Little Red Riding Hood} into an animation, introducing \toolname{}’s features along the authoring workflow.
 
\begin{figure}[t]
  \centering
  \includegraphics[width=\linewidth]{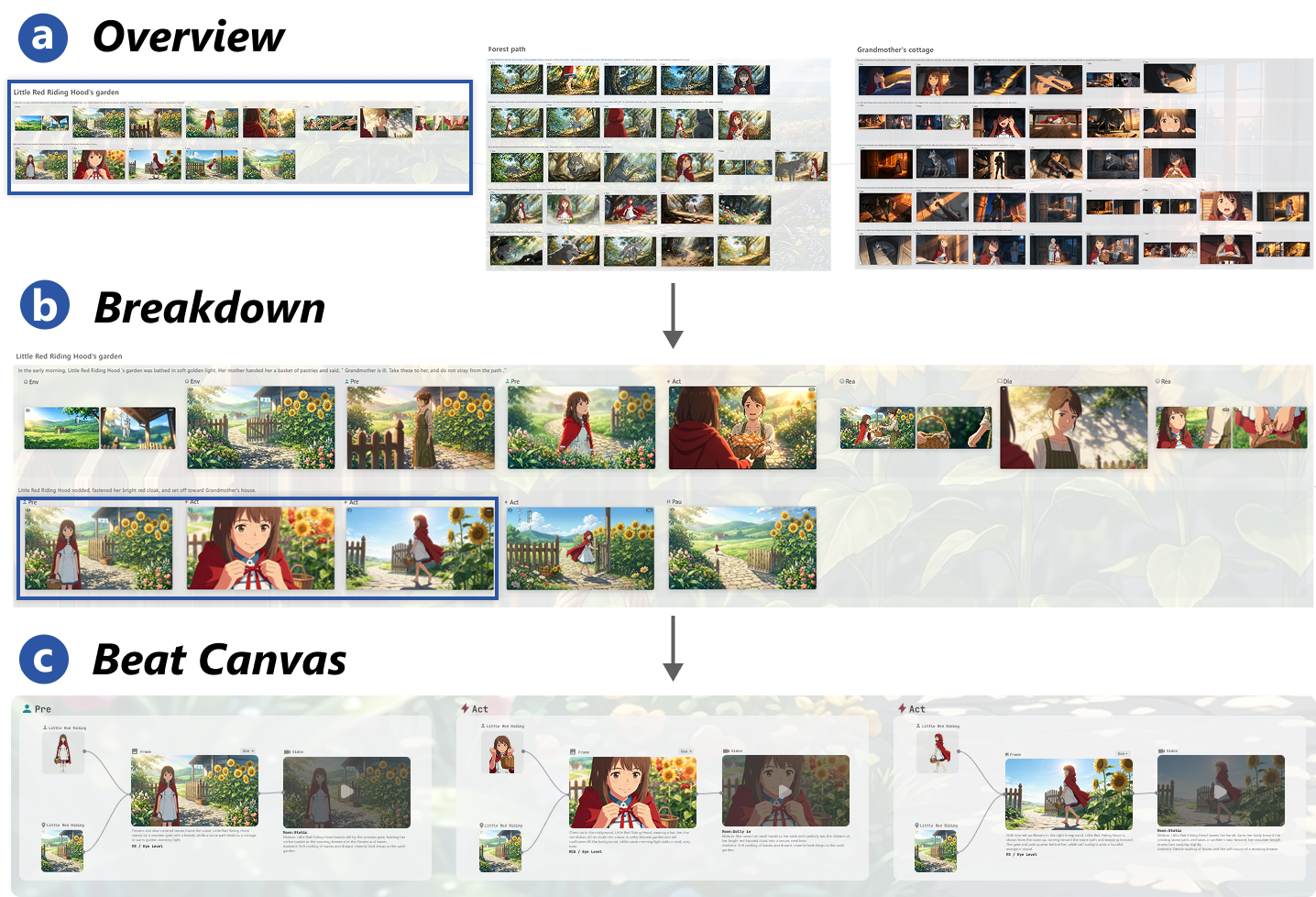}
  \caption{Progressive zoom from narrative overview to shot workspace. \iconref{a}~\overview{}: Event cards are arranged spatially by location, giving a global view of the story arc. \iconref{b}~\breakdown{}: expanding an Event reveals its Beat sequence with frame thumbnails. \iconref{c}~\beatcanvas{}: each Beat opens a local workspace where Frame and Video nodes form the shot production pipeline.}
  \label{fig:semantic-zooming}
\end{figure}
 
\subsubsection{Importing a Story.}
After the creator uploads a story text, \toolname{} automatically identifies characters (Little Red Riding Hood, Mother, Wolf, Grandmother), locations (garden, forest path, grandmother's cottage), and events, presenting them on the canvas (\autoref{fig:teaser}). The canvas arranges locations left-to-right following the story's spatial trajectory, with each location region containing its associated Event nodes. This gives the creator a global spatial overview of the narrative (DC1).
 
\subsubsection{From Event to Beat Canvas.}
Starting from an Event node, the creator first sees the system-generated Beat sequence, then progressively zooms into the shot-level workspace of a specific Beat (\autoref{fig:semantic-zooming}).

\indent\textbf{\textit{Expanding Story Events.}}
The creator selects the Event ``Little Red Riding Hood meets the Wolf in the forest.'' The system invokes the first S2S Translation to expand it into a Beat sequence: \BeatENV{} (establishing the forest atmosphere), \BeatPRE{} (the Wolf appears), \BeatDIA{} (conversation), and \BeatREA{} (Little Red Riding Hood's response) (DC2). The \scriptreader{} shows how these Beats correspond to the original story text.
 
\indent\textbf{\textit{Zooming into a Beat.}}
The creator clicks the \BeatPRE{} and the canvas zooms into the \beatcanvas{}---a local workspace for that Beat (\autoref{fig:semantic-zooming}\iconref{c}). Inside, reference frames, video clips, character assets, and location assets are arranged as nodes connected by edges that form the shot production pipeline. The second S2V Translation has already generated a set of Shot nodes with parameters, serving as a starting point for further refinement (DC2). Each Frame node defaults to \textit{Gen} mode, which preserves the system-recommended parameters and asset bindings; switching to \textit{DIY} mode clears all recommendations and lets the creator start from scratch.

\subsubsection{Crafting the Shots.}
\begin{figure*}[t]
  \centering
  \includegraphics[width=\linewidth]{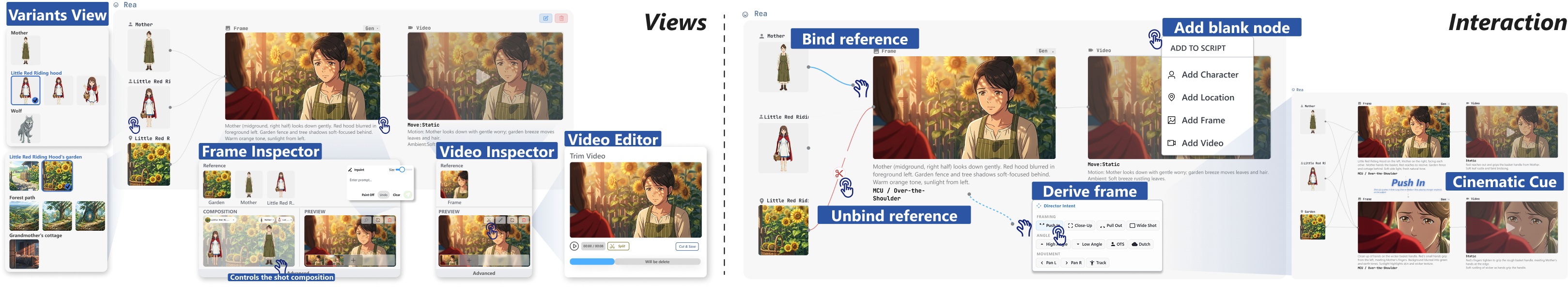}
  \caption{Designing shots in the \beatcanvas{}. \textbf{Top:} the creator inspects and edits parameters through the \frameinspector{}, \videoinspector{}, and \videoeditor{}, and browses alternatives in the \variantsview{}. \textbf{Bottom:} the workflow proceeds from binding character and location assets, to creating nodes, deriving frame variants, and pushing parameters across shots.}
  \label{fig:shot-crafting}
\end{figure*}

Once inside the \beatcanvas{}, the creator begins designing individual shots (\autoref{fig:shot-crafting}). (DC3)

\indent\textbf{\textit{Asset Binding.}}
This shot requires both Mother and Little Red Riding Hood. The creator drags suitable character variants from the \assetlib{} into the \beatcanvas{} to bind them (\autoref{fig:shot-crafting}, bottom-left). Once bound, the assets' reference images are carried into subsequent generation as visual references, helping maintain cross-shot visual consistency. Unwanted nodes can be removed via the scissor button on connecting edges.
 
\indent\textbf{\textit{Parameter Editing.}}
Selecting a Frame or Video node opens the \inspector{} panel (\autoref{fig:shot-crafting}, top-center), with controls for all nine visual parameters. The \inspector{}'s built-in \compositioneditor{} lets creators drag characters and scene elements to adjust spatial positions within the frame---for example, placing Mother at the center and placing Little Red Riding Hood's cloak in the foreground. 
 
\indent\textbf{\textit{Frame and Video Generation.}}
After confirming parameters, the system first generates a Frame (static reference image) and then, based on it, generates a Video (animated clip). When a Frame node is connected to a Video node, the generated reference image is bound to that Video as its visual reference for subsequent generation. When a creator wants to extend a Shot into a longer continuous take, the \videoinspector{} provides a continuation option that chains successive clips from the last keyframe, enabling multi-round extension into a seamless long shot while preserving visual consistency.
 
\indent\textbf{\textit{Cinematic Cues.}}
Between adjacent Shots within a Beat, blue \cinematiccue{} on the connecting edges annotate shot-to-shot visual relationships---such as shot-size continuity, axis-rule compliance, and spatial consistency---helping creators understand the system's parameter choices and spot potential issues. Creators can also drag out a custom Cue to express directorial intent (e.g. ``push to close-up''); the system then infers the corresponding parameters and generates a new Frame, making \cinematiccue{} both an explanation and an authoring channel.
 
\subsubsection{Editing and Refining.}
After initial shot design, the creator reviews the overall result and makes adjustments (\autoref{fig:refining}). Taking the passage where Mother hands the basket to Little Red Riding Hood as an example:

\begin{figure}[t]
  \centering
  \includegraphics[width=\linewidth]{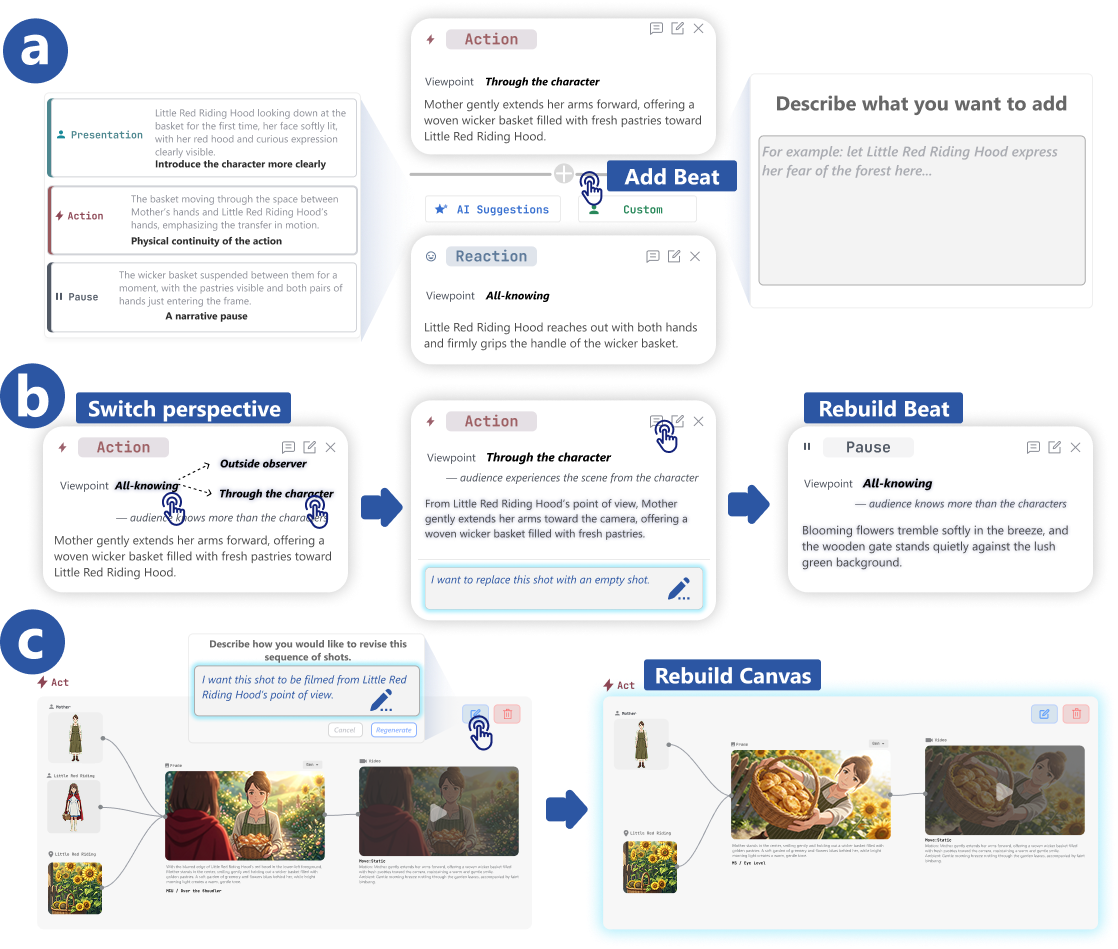}
  \caption{Two editing operations across the narrative hierarchy. \iconref{a}~\beatsuggestion{} inserts new Beats from surrounding context. \iconref{b}--\iconref{c}~Intent-based editing lets creators revise at the Beat level (\iconref{b}, e.g. viewpoint or Beat type) or Shot level (\iconref{c}, e.g. framing), with the system updating the corresponding parameters.}
  \label{fig:refining}
\end{figure}

\indent\textbf{\textit{Beat Suggestion}.}
While reviewing Beats in\breakdown{}, the creator notices that the emotional buildup before the basket handoff feels insufficient. Clicking the ``+'' icon on the connecting line between 2 adjacent Beats opens a recommendation panel (\autoref{fig:refining}\iconref{a}). The system suggests candidate Beats based on the surrounding context, each accompanied by a rationale. The creator can accept, modify, or dismiss any suggestion, or freely express an idea for how the passage should unfold, and the system then generates a corresponding Beat automatically.

\indent\textbf{\textit{Intent-based Editing}.}
When the current presentation of a moment does not feel right, the creator can directly express an intended narrative effect. At the Beat level, this may involve revising how the basket handoff is perceived---for example, changing the viewpoint from an all-knowing perspective to Little Red Riding Hood's perspective, or replacing the action Beat with a more atmospheric pause (\autoref{fig:refining}\iconref{b}). Intent-based editing also extends into the \beatcanvas{}, where the creator can describe how a local shot should be revised---for example, requesting that the shot be filmed from Little Red Riding Hood's point of view. The system then updates the corresponding shot realization based on the expressed intent (\autoref{fig:refining}\iconref{c}).

\indent\textbf{\textit{Multi-level Editing}.}
Creators can edit results at the Event, Beat, and Shot levels. When an upstream level is modified, the system marks the affected downstream results on the canvas and allows the creator to selectively regenerate them. Downstream edits made by the creator are preserved as explicit overrides and are not automatically overwritten by later upstream changes.
\subsection{Two-Stage Translation Pipeline}
\label{sec:generation-text}

\toolname{} instantiates the translation rules in \S3.4 as a two-stage authoring pipeline (\autoref{fig:pipeline}). Rather than directly prompting a video model from raw story text, the system first performs S2S (\autoref{fig:pipeline}\iconref{a}), which expands an Event into an editable Beat sequence shown in the \breakdown{} and aligned in the \scriptreader{} for revision before shot authoring. It then performs S2V (\autoref{fig:pipeline}\iconref{b}), which realizes each Beat as editable Shot structures with initial visual parameters and instantiates them in the \beatcanvas{} as Frame and Video nodes. Generation then proceeds from a static Frame to a Video clip, exposing intermediate narrative and visual decisions for interactive editing rather than treating story-to-video generation as a single end-to-end step.

\subsection{Implementation}
 \toolname{} is built with Vue~3 and TypeScript on the front end and FastAPI and Python on the back end. The canvas uses Vue Flow~\cite{vueflow2025}, the \compositioneditor{} in the \inspector{} uses PixiJS~\cite{pixijs}, and the cross-layer alignment in the \scriptreader{} is drawn with D3~\cite{bostock2011d3}. The LLM interface is configurable and this study uses Google Gemini (gemini-3-pro)~\cite{gemini}. Images are generated with Seedream~4.5 and Gemini~3.1 Flash Image~\cite{seedream2025seedream, gemini}, videos with Veo~3 Generate~\cite{google2025veo3}, and frame interpolation with RIFE~\cite{huang2022real}.

\section{Evaluation}
Our evaluation is organized around 3 research questions:

\textbf{RQ1.} How effectively does \toolname{} support creativity and authoring experience for everyday creators?

\textbf{RQ2.} How do creators with different levels of experience use \toolname{}'s design framework and interaction mechanisms to shape their narrative and cinematic intent?

\textbf{RQ3.} To what extent do the three-layer design framework and its translation rules align with professional filmmaking practice, and where are their boundaries?

To answer these questions, we conducted 3 complementary studies: (1) a within-subjects user study comparing \toolname{} with a Baseline condition in terms of creativity support and authoring experience (RQ1); (2) 2 exploratory case studies tracing the full creative process of creators with different backgrounds (RQ2); and (3) expert interviews examining the framework and outputs from a professional filmmaking perspective (RQ3).
\subsection{User Study}
To answer \textbf{RQ1}, we conducted a within-subjects user study comparing \toolname{} with a Baseline condition in terms of creativity support and authoring experience.

\textbf{Participants and Task.}
We recruited 16 participants with no filmmaking background (8 female, 8 male, aged 20--26, $M=22.8$).
In a within-subjects design, each participant used both \toolname{} and a Baseline condition to create a complete animated video from a short story.
The Baseline provided the same underlying generative model as \toolname{} (i.e., identical generation capabilities), but did not provide \toolname{}'s structured intermediate representations, semantic zooming, or rule-guided interaction mechanisms. Participants in the Baseline condition could freely converse with the LLM for narrative planning and creative iteration, and could use general-purpose external tools of their choice to arrange visuals, organize ideas, and assemble clips. This setup allowed us to compare generative capability with and without \toolname{}'s structured authoring support.
We used 2 fairy tales of comparable complexity, \textit{Alice in Wonderland} and \textit{Little Red Riding Hood}, each approximately 170 words. Stories were counterbalanced across conditions and participants using a Latin square design (4 groups $\times$ 4 participants).

\textbf{Study Procedure.}
Each participant completed one session per condition, with each session lasting approximately 2 hours (roughly 4 hours total across both conditions).
In each session, participants first received a tutorial on the assigned tool and then created a full animated video from the assigned story.
After completing the task, they took part in a semi-structured interview reflecting on their creative process and experience.
Cognitive load was measured using the Raw NASA-TLX~\cite{hart2006nasa}, and creativity support was measured using the CSI~\cite{cherry2014quantifying}. Because this is a single user authoring task, we report the 5 applicable CSI factors and exclude Collaboration. We also collected a post-test questionnaire containing 9 bipolar preference items (scored from $-10$ to $+10$) across 4 dimensions: \textit{Authoring Experience} (workflow naturalness and interaction clarity; Q1--Q2), \textit{Creative Control} (exploration flexibility and creative agency; Q3--Q4), \textit{Intent Expression} (intent-to-visual accuracy and cinematic guidance; Q5--Q6), and \textit{Overall Assessment} (output quality, tool preference, and creative potential; Q7--Q9).
Pairwise comparisons used the Wilcoxon signed-rank test, with matched-pairs rank-biserial correlation $r$ reported as the effect size.

\begin{figure*}[t]
    \centering
    \includegraphics[width=\textwidth]{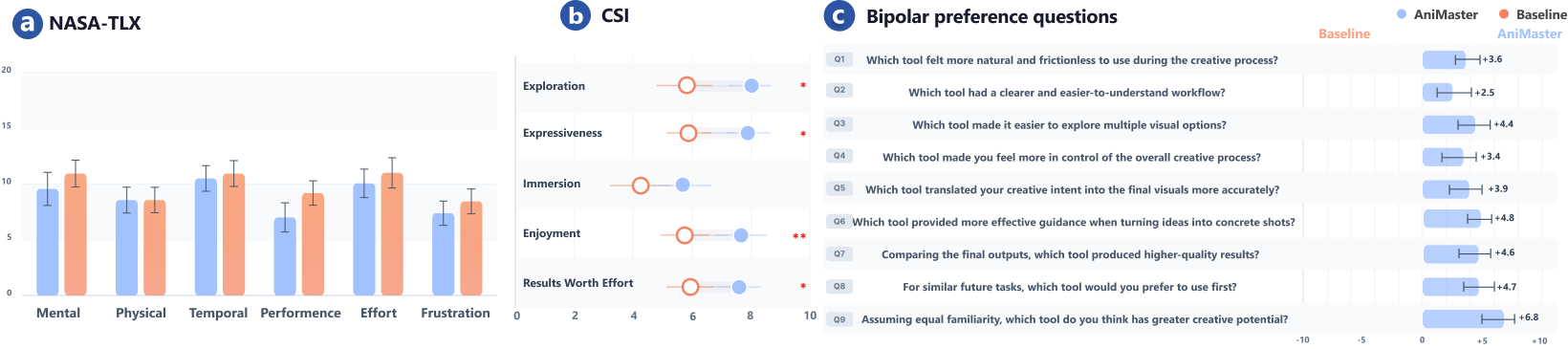}
    \caption{Quantitative results ($N=16$, within-subjects). \iconref{a}~NASA-TLX subscale scores. \iconref{b}~CSI five-factor paired comparison. \iconref{c}~9 bipolar preference items.}
    \label{fig:user-study}
\end{figure*}

\textbf{Quantitative results.}
\toolname{} scored significantly higher on overall CSI ($7.16$ vs.\ $5.70$, $p=.012$, $r=.71$). Four of five factors also showed significant differences (all $r>.62$), except Immersion ($p=.211$). NASA-TLX total scores did not differ significantly ($p=.597$). All 9 bipolar preference items favored \toolname{} (all $p<.05$), with Creative Potential and Shot Guidance Effectiveness rated highest. All 16 participants preferred \toolname{}. Together, these findings suggest that \toolname{} improved creativity support, perceived guidance, and overall authoring experience without increasing total task load.

\textbf{Qualitative results.}
Interview recordings were transcribed and open-coded independently by 2 authors, who then reconciled discrepancies through inductive thematic analysis~\cite{braun2006using}. Initial inter-coder agreement reached Cohen's $\kappa=.74$ before reconciliation. Two recurring themes emerged.

\begin{enumerate}[label=\textbf{(\arabic*)}, leftmargin=*]
    \item \textbf{AI as collaborator.} Participants generally perceived the AI in \toolname{} as a collaborator rather than a passive executor. As \parti{P7} noted, \textit{``It felt more like an experienced partner---I just needed to describe what happens in the story, and it could break it down into shots from a professional perspective.''} This reflects a shift from prompt engineer to creative director.

    \item \textbf{Framework as scaffolding.} The design framework served as professional scaffolding for participants without a filmmaking background. \parti{P4} explained, \textit{``It was like thinking as a director---splitting the story into storyboard segments, then deciding how to shoot each one, including fairly technical terms like foreground, midground, and background.''}
\end{enumerate}

\begin{figure}[t]
    \centering
    \includegraphics[width=\linewidth]{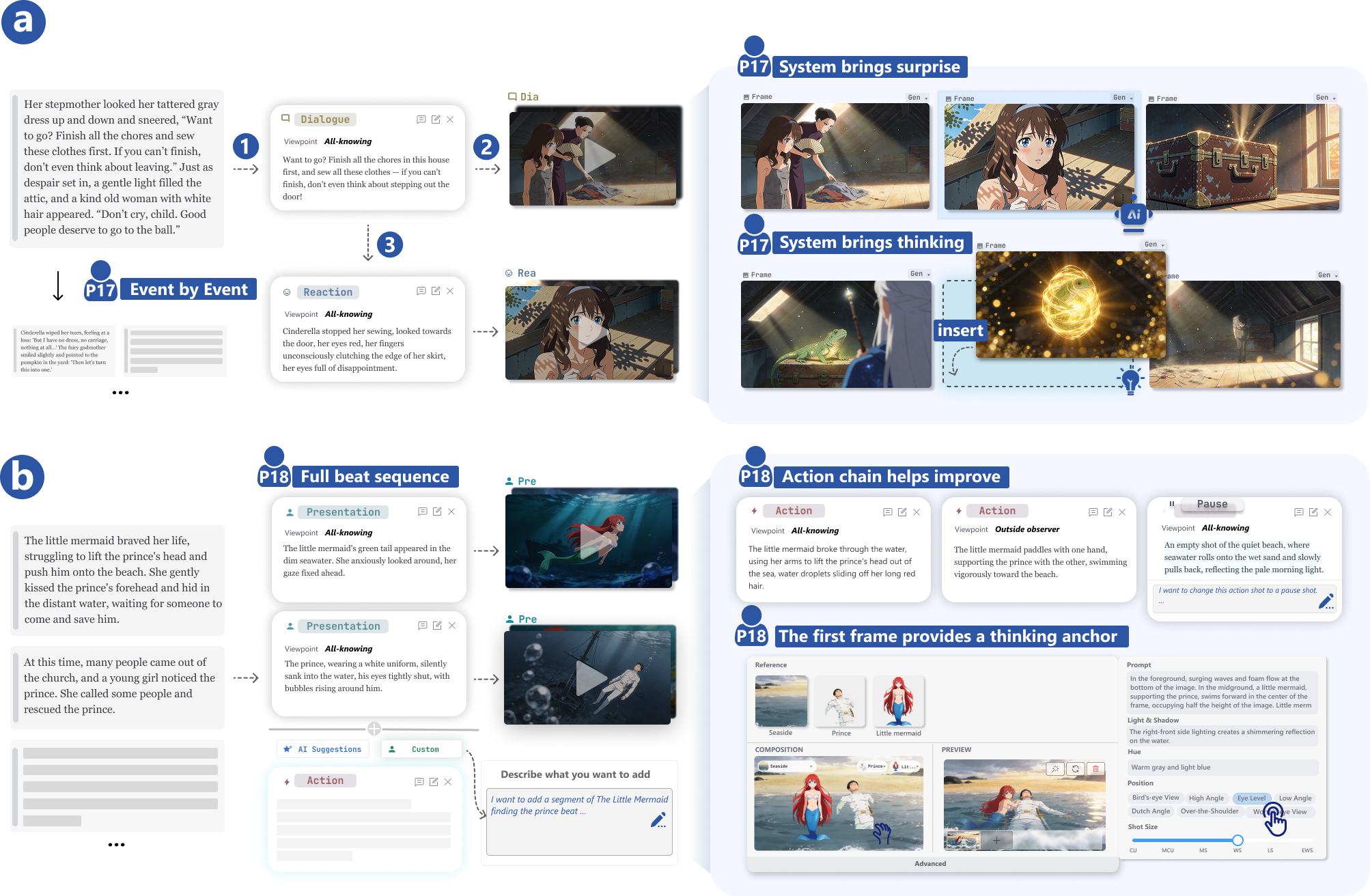}
    \caption{Case study overview. 2 participants created animated videos using \toolname{}: \parti{P17} adapted \textit{Cinderella} and \parti{P18} adapted \textit{The Little Mermaid}.}
    \label{fig:case-study}
\end{figure}

\subsection{Case Study}

To answer \textbf{RQ2}, we conducted 2 exploratory case studies to examine how creators with different backgrounds used \toolname{}'s design framework and interaction mechanisms throughout a full authoring process.

\textbf{Participants and Procedure.}
We invited 2 new participants for authoring sessions. \parti{P17}, a college student with no filmmaking experience, chose \textit{Cinderella}. \parti{P18}, a hobbyist with 2 years of animation experience, chose \textit{The Little Mermaid}. There was no time limit. We tracked the process using think-aloud protocols, screen recordings, and retrospective interviews based on video playback.

\textbf{Case Findings.}

\textit{\parti{P17}: From Following Suggestions to Active Judgments.}
\parti{P17} adopted an event-by-event strategy and initially relied on the system-generated Beat sequence and suggested shots, because many were ones he ``would never have thought of on his own.''

A turning point came when the system generated a reaction shot of Cinderella with tears in her eyes: \textit{``I never imagined this shot could be done this way.''} This helped him realize that shots can carry emotional expression, not just depict events. He then began making revisions---for example, requesting an establishing shot when a lizard-turned-butler appeared too abruptly.

After completing all Events, \parti{P17} shifted from ``following the workflow'' to ``making local revisions based on coherence,'' preferring high-level intent edits because he ``had ideas but could not express them as shot parameters.'' He described \toolname{} as a \textit{``teacher''} that guided him in learning to tell a story through shots.

\textit{\parti{P18}: Global Planning and Deliberate Trade-offs.}
Unlike \parti{P17}, \parti{P18} first expanded the story into a Beat sequence, treating it as a reviewable structure before moving to shot-level authoring. He used \beatsuggestion{} to fill in weak spots---for instance, adding Action Beats to the rescue sequence where the mermaid drags the unconscious prince ashore, because the action chain jumped too abruptly from struggle to arrival on the beach (\autoref{fig:case-study}\iconref{b}).

In the \beatcanvas{}, \parti{P18} used relationship cues between adjacent shots to assess action progression and refine the rescue sequence into a coherent action chain. He also relied on system-generated first frames as ``thinking anchors'' for composition---adjusting character placement and camera angles based on the visual reference before committing to full generation. He described \toolname{} as an \textit{``assistant''}: \textit{``The generated templates are great suggestions, but when it comes to the actual creation, I still rely on my own style.''}

\textbf{Cross-Case Observation.}
The 2 cases reveal distinct creative paths. \parti{P17} used \toolname{} as scaffolding, gradually moving from accepting system suggestions to making deliberate shot-level judgments. \parti{P18} used it as an assistant that helped organize his intuitions through Beat-level planning and local refinement. Despite these different strategies, both participants produced complete works in about 4 hours, suggesting that \toolname{} can support creators with different experience levels without requiring a mode switch.

\subsection{Expert Interview}

\textbf{Participants and Procedure.}
To evaluate \toolname{}'s alignment with professional practice (\textbf{RQ3}), we interviewed two experts: \parti{E7}, an animation director (8 years exp.), and \parti{E8}, a university educator and practitioner (15 years exp.). After exploring \toolname{} and reviewing case studies from \parti{P17} and \parti{P18}, each participated in a 70–80 minute semi-structured interview. These sessions focused on shot logic, narrative pacing, and workflow alignment.

\textbf{Professional validity of shot-level narrative logic.} Both experts confirmed that the system-generated shots served distinct narrative functions rather than merely illustrating plot events. Regarding \parti{P17}'s \textit{Cinderella}, \parti{E7} affirmed the clarity of its single-scene causal progression: \textit{``The causal chain in the magic transformation sequence is complete---it first establishes the oppressive attic space, then introduces the fairy godmother, and uses the reveals of the gown and glass slipper to drive the narrative. The shots are not just pretty images; each one serves a narrative function.''} Regarding \parti{P18}'s \textit{The Little Mermaid}, \parti{E8} noted a stronger sense of cinematic quality: \textit{``The wide shots of the sea, the isolated figure, and the storm establish a fairly clear emotional arc. The system is not just generating events---it is attempting to generate atmosphere for entire passages.''}

\textbf{E7} observed that the system differentiates among narrative roles---establishing mood, introducing characters, advancing action, and capturing reactions---within its shot generation. The presence of \BeatPAU{} was particularly noteworthy: \textit{``Pause is an important pacing device. Most AI systems skip this step entirely.''} Together, these comments suggest that the Beat-type taxonomy captures meaningful professional distinctions in how shots function narratively.

\textbf{Alignment with professional workflows and stylistic scope.}
When reviewing the authoring canvas, both experts recognized a structural correspondence between the system's three-layer workflow and their own practice. \parti{E7} remarked, \textit{``When I get a literary script, I first do narrative segmentation, then create a storyboard, and finally move into production. You have made this process explicit, so that non-professionals can walk through the same path.''}

Both experts also valued Focalization as an independently adjustable dimension that affects emotional focus and information distribution. At the same time, \parti{E7} noted that the underlying translation rules lean toward classical continuity editing: \textit{``It provides a reliable classical narrative baseline, but it is still conservative for more experimental forms of expression.''} This suggests that the authoring logic is most effective for plot-driven, continuity-based narratives.

\textbf{Overall Assessment.}
Overall, the expert interviews suggest that \toolname{}'s authoring logic corresponds to professional workflows and that both creative paths observed in \S5.2 are valid ways of working with the system. At the same time, the interviews reveal a gap between local reliability and long-range coherence. The system is best suited for classical-narrative-oriented previsualization, while stylistic innovation, cross-scene transitions, and cross-shot consistency remain open areas for improvement.

\section{Discussion}
 We synthesize findings from the user study, case studies, and expert review, and discuss implications, limitations, and future directions.
 
\subsection{General Discussion}
The user study, case studies, and expert interviews converge on several observations about how \toolname{}'s design framework shapes the authoring experience. Rather than repeating individual findings, we focus here on three cross-cutting themes that emerged when synthesizing across the studies.
 
\textbf{Professional Knowledge Scaffolding.} 
Across the three studies, we found that \toolname{}'s intermediate representations---Beats, Shots, and their visual parameters---helped participants engage with filmmaking concepts that would otherwise remain implicit. Participants in the user study described the AI as a collaborator that could break stories down into professional shot logic, while the case studies showed that the same structure could function as either a ``teacher'' or an ``assistant'' depending on the creator's experience. The experts similarly noted that \toolname{} makes professional decision steps explicit and editable. Together, these findings suggest that visible intermediate representations may be more valuable here than end-to-end generation alone.
 
\textbf{Cognitive Load Redistribution.}
The non-significant difference in NASA-TLX overall scores ($p = .597$) suggests that \toolname{} did not reduce total cognitive effort. Instead, \toolname{} appears to redirect users' effort toward a different \textit{type} of cognitive work. In the Baseline condition, participants spent considerable effort on low-level prompt iteration; in \toolname{}, their attention appeared to shift toward higher-level decisions about shot logic and narrative pacing. \parti{P17's} trajectory in the case study reflects this pattern: after initially accepting system outputs passively, \parti{P17} began actively modifying shots once he recognized that ``a shot can carry emotion''---a progression consistent with the extraneous-to-germane load shift described in Cognitive Load Theory~\cite{sweller1988cognitive}. For creative authoring tools, this suggests that an important design goal may be not simply to minimize workload, but to channel users' effort toward meaningful creative judgment.
 
\textbf{Narrative-Structured Free-Form Authoring.}
When creators entered the \beatcanvas{}, we observed a shift from linear, step-by-step operation to spatial, exploratory creation. The \maincanvas{}'s three semantic zoom levels---\overview{}, \breakdown{}, and \beatcanvas{}---scope the range of attention at each level of the hierarchy, allowing free exploration within each layer while the narrative structure maintains global coherence. In the case studies, \parti{P18} freely rearranged Beat sequences and adjusted camera angles in the \beatcanvas{} while the three-layer structure kept his edits narratively coherent. This design supports a continuum from guided to autonomous creation without requiring explicit mode switching.
\subsection{Limitations and Future Work}
While results are encouraging, interviews and our analysis reveal two key gaps. Both suggest future directions.
 
\textbf{Theoretical Scope of Translation.}
The Translation rules are grounded in classical Hollywood continuity editing. As \parti{E7} noted, the rules ``provide a reliable classical narrative baseline but remain conservative for more experimental expression.'' Support for nonlinear narrative structures (e.g. flashbacks, parallel montage) remains limited. Additionally, Translation is currently a uniform static prompt and does not adapt to creators; future work could introduce timeline branching in the Story Space and evolve Translation into interactive guidance that adjusts based on operation history.
 
\textbf{Cross-Scene Narrative Planning.}
The system currently expands shots Event-by-Event in linear order. Within-scene orchestration is relatively complete, but cross-scene techniques such as foreshadowing and callbacks are absent. \parti{E8} observed that some results ``resemble a concatenation of high-quality illustrations rather than a sequence driven by continuous shot logic.'' The controlled experiment used short stories of approximately 170 words; scalability to longer narratives with more complex cross-scene dependencies remains to be validated. Future work should introduce cross-Event narrative arc planning and parallel temporal tracks.

\textbf{Generation-layer constraints.}
Even when AniMaster produces a coherent Beat sequence and shot plan, the final outputs may still break down at the level of cross-shot visual consistency and scene-to-scene continuity. This gap suggests that the current system is better understood as a structured authoring and previsualization scaffold than as a fully reliable end-to-end production pipeline. Future work should therefore combine our framework with stronger sequence-level consistency mechanisms in video generation.

\section{Conclusion}

We present \toolname{}, an LLM-powered authoring tool that helps everyday creators produce cinematic animated videos from free-form story texts. Drawing on narratology and film studies, we propose a three-layer design framework that organizes the key dimensions across Story, Script, and Video, and defines the Translation rules connecting them. \toolname{} instantiates this framework through rule-based translation and a canvas-based interface with semantic zooming, letting creators review, modify, and override system decisions at every layer. The structured intermediate representations---Events, Beats, and Shots---simultaneously serve as a cognitive scaffold for novices and an organizational skeleton for experienced creators, supporting a continuum from guided to autonomous authoring. A user study ($N = 16$), two case studies, and expert interviews demonstrate that \toolname{} significantly improves creativity support and authoring experience. We hope this approach---encoding tacit professional knowledge as editable, structured representations---can extend to other creative domains beyond cinematic animation.

\begin{acks}
To Robert, for the bagels and explaining CMYK and color spaces.
\end{acks}

\bibliographystyle{ACM-Reference-Format}
\bibliography{sample-base}


\begin{thebibliography}{57}


\ifx \showCODEN    \undefined \def \showCODEN     #1{\unskip}     \fi
\ifx \showISBNx    \undefined \def \showISBNx     #1{\unskip}     \fi
\ifx \showISBNxiii \undefined \def \showISBNxiii  #1{\unskip}     \fi
\ifx \showISSN     \undefined \def \showISSN      #1{\unskip}     \fi
\ifx \showLCCN     \undefined \def \showLCCN      #1{\unskip}     \fi
\ifx \shownote     \undefined \def \shownote      #1{#1}          \fi
\ifx \showarticletitle \undefined \def \showarticletitle #1{#1}   \fi
\ifx \showURL      \undefined \def \showURL       {\relax}        \fi
\providecommand\bibfield[2]{#2}
\providecommand\bibinfo[2]{#2}
\providecommand\natexlab[1]{#1}
\providecommand\showeprint[2][]{arXiv:#2}

\bibitem[Avrahami et~al\mbox{.}(2024)]%
        {avrahami2024chosen}
\bibfield{author}{\bibinfo{person}{Omri Avrahami}, \bibinfo{person}{Amir
  Hertz}, \bibinfo{person}{Yael Vinker}, \bibinfo{person}{Moab Arar},
  \bibinfo{person}{Shlomi Fruchter}, \bibinfo{person}{Ohad Fried},
  \bibinfo{person}{Daniel Cohen-Or}, {and} \bibinfo{person}{Dani Lischinski}.}
  \bibinfo{year}{2024}\natexlab{}.
\newblock \showarticletitle{The chosen one: Consistent characters in
  text-to-image diffusion models}. In \bibinfo{booktitle}{\emph{ACM SIGGRAPH
  2024 conference papers}}. \bibinfo{pages}{1--12}.
\newblock


\bibitem[Bal(2017)]%
        {bal2017narratology}
\bibfield{author}{\bibinfo{person}{Mieke Bal}.}
  \bibinfo{year}{2017}\natexlab{}.
\newblock \bibinfo{booktitle}{\emph{Narratology: Introduction to the Theory of
  Narrative}}.
\newblock \bibinfo{publisher}{University of Toronto Press},
  \bibinfo{address}{Toronto}.
\newblock


\bibitem[Barthes(1966)]%
        {barthes1966introduction}
\bibfield{author}{\bibinfo{person}{Roland Barthes}.}
  \bibinfo{year}{1966}\natexlab{}.
\newblock \showarticletitle{Introduction {\`a} l'analyse structurale des
  r{\'e}cits}.
\newblock \bibinfo{journal}{\emph{Communications}} \bibinfo{volume}{8},
  \bibinfo{number}{1} (\bibinfo{year}{1966}), \bibinfo{pages}{1--27}.
\newblock


\bibitem[Blattmann et~al\mbox{.}(2023)]%
        {blattmann2023stable}
\bibfield{author}{\bibinfo{person}{Andreas Blattmann}, \bibinfo{person}{Tim
  Dockhorn}, \bibinfo{person}{Sumith Kulal}, \bibinfo{person}{Daniel
  Mendelevitch}, \bibinfo{person}{Maciej Kilian}, \bibinfo{person}{Dominik
  Lorenz}, \bibinfo{person}{Yam Levi}, \bibinfo{person}{Zion English},
  \bibinfo{person}{Vikram Voleti}, \bibinfo{person}{Adam Letts},
  {et~al\mbox{.}}} \bibinfo{year}{2023}\natexlab{}.
\newblock \showarticletitle{Stable video diffusion: Scaling latent video
  diffusion models to large datasets}.
\newblock \bibinfo{journal}{\emph{arXiv preprint arXiv:2311.15127}}
  (\bibinfo{year}{2023}).
\newblock


\bibitem[Bordwell(1985)]%
        {bordwell1985narration}
\bibfield{author}{\bibinfo{person}{David Bordwell}.}
  \bibinfo{year}{1985}\natexlab{}.
\newblock \bibinfo{booktitle}{\emph{Narration in the Fiction Film}}.
\newblock \bibinfo{publisher}{University of Wisconsin Press},
  \bibinfo{address}{Madison, WI}.
\newblock
\showISBNx{9780299101749}


\bibitem[Bordwell(2006)]%
        {bordwell2006way}
\bibfield{author}{\bibinfo{person}{David Bordwell}.}
  \bibinfo{year}{2006}\natexlab{}.
\newblock \bibinfo{booktitle}{\emph{The way Hollywood tells it: Story and style
  in modern movies}}.
\newblock \bibinfo{publisher}{Univ of California Press}.
\newblock


\bibitem[Bordwell et~al\mbox{.}(2019)]%
        {bordwell2019film}
\bibfield{author}{\bibinfo{person}{David Bordwell}, \bibinfo{person}{Kristin
  Thompson}, {and} \bibinfo{person}{Jeff Smith}.}
  \bibinfo{year}{2019}\natexlab{}.
\newblock \bibinfo{booktitle}{\emph{Film Art: An Introduction}
  (\bibinfo{edition}{12th} ed.)}.
\newblock \bibinfo{publisher}{McGraw-Hill Education}.
\newblock


\bibitem[Bostock et~al\mbox{.}(2011)]%
        {bostock2011d3}
\bibfield{author}{\bibinfo{person}{Michael Bostock}, \bibinfo{person}{Vadim
  Ogievetsky}, {and} \bibinfo{person}{Jeffrey Heer}.}
  \bibinfo{year}{2011}\natexlab{}.
\newblock \showarticletitle{D$^3$ data-driven documents}.
\newblock \bibinfo{journal}{\emph{IEEE transactions on visualization and
  computer graphics}} \bibinfo{volume}{17}, \bibinfo{number}{12}
  (\bibinfo{year}{2011}), \bibinfo{pages}{2301--2309}.
\newblock


\bibitem[Brade et~al\mbox{.}(2023)]%
        {brade2023promptify}
\bibfield{author}{\bibinfo{person}{Stephen Brade}, \bibinfo{person}{Bryan
  Wang}, \bibinfo{person}{Mauricio Sousa}, \bibinfo{person}{Sageev Oore}, {and}
  \bibinfo{person}{Tovi Grossman}.} \bibinfo{year}{2023}\natexlab{}.
\newblock \showarticletitle{Promptify: Text-to-image generation through
  interactive prompt exploration with large language models}. In
  \bibinfo{booktitle}{\emph{Proceedings of the 36th Annual ACM Symposium on
  User Interface Software and Technology}}. \bibinfo{pages}{1--14}.
\newblock


\bibitem[Braun and Clarke(2006)]%
        {braun2006using}
\bibfield{author}{\bibinfo{person}{Virginia Braun} {and}
  \bibinfo{person}{Victoria Clarke}.} \bibinfo{year}{2006}\natexlab{}.
\newblock \showarticletitle{Using thematic analysis in psychology}.
\newblock \bibinfo{journal}{\emph{Qualitative research in psychology}}
  \bibinfo{volume}{3}, \bibinfo{number}{2} (\bibinfo{year}{2006}),
  \bibinfo{pages}{77--101}.
\newblock


\bibitem[{ByteDance Seed Team}(2026)]%
        {bytedance2026seedance2}
\bibfield{author}{\bibinfo{person}{{ByteDance Seed Team}}.}
  \bibinfo{year}{2026}\natexlab{}.
\newblock \bibinfo{booktitle}{\emph{Seed2.0 Model Card: Towards Intelligence
  Frontier for Real-World Complexity}}.
\newblock \bibinfo{type}{{T}echnical {R}eport}.
  \bibinfo{institution}{ByteDance}.
\newblock
\urldef\tempurl%
\url{https://seed.bytedance.com/en/seed2}
\showURL{%
\tempurl}
\newblock
\shownote{Technical Report}.


\bibitem[Cakmakoglu(2025)]%
        {vueflow2025}
\bibfield{author}{\bibinfo{person}{Burak Cakmakoglu}.}
  \bibinfo{year}{2025}\natexlab{}.
\newblock \bibinfo{booktitle}{\emph{Vue Flow: A highly customizable Vue 3
  Flowchart component}}.
\newblock
\urldef\tempurl%
\url{https://github.com/bcakmakoglu/vue-flow}
\showURL{%
\tempurl}


\bibitem[Cao et~al\mbox{.}(2025)]%
        {cao2025compositional}
\bibfield{author}{\bibinfo{person}{Yining Cao}, \bibinfo{person}{Yiyi Huang},
  \bibinfo{person}{Anh Truong}, \bibinfo{person}{Hijung~Valentina Shin}, {and}
  \bibinfo{person}{Haijun Xia}.} \bibinfo{year}{2025}\natexlab{}.
\newblock \showarticletitle{Compositional structures as substrates for human-ai
  co-creation environment: A design approach and a case study}. In
  \bibinfo{booktitle}{\emph{Proceedings of the 2025 CHI Conference on Human
  Factors in Computing Systems}}. \bibinfo{pages}{1--25}.
\newblock


\bibitem[Chatman and Chatman(1978)]%
        {chatman1978story}
\bibfield{author}{\bibinfo{person}{Seymour~Benjamin Chatman} {and}
  \bibinfo{person}{Seymour Chatman}.} \bibinfo{year}{1978}\natexlab{}.
\newblock \bibinfo{booktitle}{\emph{Story and discourse: Narrative structure in
  fiction and film}}.
\newblock \bibinfo{publisher}{Cornell university press}.
\newblock


\bibitem[Chatterjee et~al\mbox{.}(2025)]%
        {chatterjee2025stable}
\bibfield{author}{\bibinfo{person}{Agneet Chatterjee}, \bibinfo{person}{Rahim
  Entezari}, \bibinfo{person}{Maksym Zhuravinskyi}, \bibinfo{person}{Maksim
  Lapin}, \bibinfo{person}{Reshinth Adithyan}, \bibinfo{person}{Amit Raj},
  \bibinfo{person}{Chitta Baral}, \bibinfo{person}{Yezhou Yang}, {and}
  \bibinfo{person}{Varun Jampani}.} \bibinfo{year}{2025}\natexlab{}.
\newblock \showarticletitle{Stable Cinemetrics: Structured Taxonomy and
  Evaluation for Professional Video Generation}.
\newblock \bibinfo{journal}{\emph{arXiv preprint arXiv:2509.26555}}
  (\bibinfo{year}{2025}).
\newblock


\bibitem[Chen et~al\mbox{.}(2024)]%
        {chen2024cinepregen}
\bibfield{author}{\bibinfo{person}{Yiran Chen}, \bibinfo{person}{Anyi Rao},
  \bibinfo{person}{Xuekun Jiang}, \bibinfo{person}{Shishi Xiao},
  \bibinfo{person}{Ruiqing Ma}, \bibinfo{person}{Zeyu Wang},
  \bibinfo{person}{Hui Xiong}, {and} \bibinfo{person}{Bo Dai}.}
  \bibinfo{year}{2024}\natexlab{}.
\newblock \showarticletitle{Cinepregen: Camera controllable video
  previsualization via engine-powered diffusion}.
\newblock \bibinfo{journal}{\emph{arXiv preprint arXiv:2408.17424}}
  (\bibinfo{year}{2024}).
\newblock


\bibitem[Cherry and Latulipe(2014)]%
        {cherry2014quantifying}
\bibfield{author}{\bibinfo{person}{Erin Cherry} {and} \bibinfo{person}{Celine
  Latulipe}.} \bibinfo{year}{2014}\natexlab{}.
\newblock \showarticletitle{Quantifying the creativity support of digital tools
  through the creativity support index}.
\newblock \bibinfo{journal}{\emph{ACM Transactions on Computer-Human
  Interaction (TOCHI)}} \bibinfo{volume}{21}, \bibinfo{number}{4}
  (\bibinfo{year}{2014}), \bibinfo{pages}{1--25}.
\newblock


\bibitem[Chung and Kreminski(2024)]%
        {chung2024patchview}
\bibfield{author}{\bibinfo{person}{John Joon~Young Chung} {and}
  \bibinfo{person}{Max Kreminski}.} \bibinfo{year}{2024}\natexlab{}.
\newblock \showarticletitle{Patchview: Llm-powered worldbuilding with
  generative dust and magnet visualization}. In
  \bibinfo{booktitle}{\emph{Proceedings of the 37th Annual ACM Symposium on
  User Interface Software and Technology}}. \bibinfo{pages}{1--19}.
\newblock


\bibitem[DeepMind(2025)]%
        {google2025veo3}
\bibfield{author}{\bibinfo{person}{Google DeepMind}.}
  \bibinfo{year}{2025}\natexlab{}.
\newblock \bibinfo{booktitle}{\emph{Veo: a text-to-video generation system}}.
\newblock \bibinfo{type}{{T}echnical {R}eport}. \bibinfo{institution}{Google
  DeepMind}.
\newblock
\urldef\tempurl%
\url{https://storage.googleapis.com/deepmind-media/veo/Veo-3-Tech-Report.pdf}
\showURL{%
\tempurl}


\bibitem[Elmoghany et~al\mbox{.}(2025)]%
        {elmoghany2025survey}
\bibfield{author}{\bibinfo{person}{Mohamed Elmoghany}, \bibinfo{person}{Ryan
  Rossi}, \bibinfo{person}{Seunghyun Yoon}, \bibinfo{person}{Subhojyoti
  Mukherjee}, \bibinfo{person}{Eslam~Mohamed Bakr}, \bibinfo{person}{Puneet
  Mathur}, \bibinfo{person}{Gang Wu}, \bibinfo{person}{Viet~Dac Lai},
  \bibinfo{person}{Nedim Lipka}, \bibinfo{person}{Ruiyi Zhang},
  {et~al\mbox{.}}} \bibinfo{year}{2025}\natexlab{}.
\newblock \showarticletitle{A survey on long-video storytelling generation:
  architectures, consistency, and cinematic quality}. In
  \bibinfo{booktitle}{\emph{Proceedings of the IEEE/CVF International
  Conference on Computer Vision}}. \bibinfo{pages}{7023--7035}.
\newblock


\bibitem[Genette(1980)]%
        {genette1980narrative}
\bibfield{author}{\bibinfo{person}{G{\'e}rard Genette}.}
  \bibinfo{year}{1980}\natexlab{}.
\newblock \bibinfo{booktitle}{\emph{Narrative discourse: An essay in method}}.
  Vol.~\bibinfo{volume}{3}.
\newblock \bibinfo{publisher}{Cornell University Press}.
\newblock


\bibitem[{Google DeepMind}(2025)]%
        {gemini}
\bibfield{author}{\bibinfo{person}{{Google DeepMind}}.}
  \bibinfo{year}{2025}\natexlab{}.
\newblock \bibinfo{title}{Gemini}.
\newblock
\urldef\tempurl%
\url{https://deepmind.google/models/gemini/}
\showURL{%
\tempurl}


\bibitem[Groves and {PixiJS Team}({[n.\,d.]})]%
        {pixijs}
\bibfield{author}{\bibinfo{person}{Mat Groves} {and} \bibinfo{person}{{PixiJS
  Team}}.} \bibinfo{year}{[n.\,d.]}\natexlab{}.
\newblock \bibinfo{booktitle}{\emph{PixiJS}}.
\newblock
\urldef\tempurl%
\url{https://github.com/pixijs/pixijs}
\showURL{%
\tempurl}


\bibitem[Hart(2006)]%
        {hart2006nasa}
\bibfield{author}{\bibinfo{person}{Sandra~G Hart}.}
  \bibinfo{year}{2006}\natexlab{}.
\newblock \showarticletitle{NASA-task load index (NASA-TLX); 20 years later}.
  In \bibinfo{booktitle}{\emph{Proceedings of the human factors and ergonomics
  society annual meeting}}, Vol.~\bibinfo{volume}{50}. Sage publications Sage
  CA: Los Angeles, CA, \bibinfo{pages}{904--908}.
\newblock


\bibitem[He et~al\mbox{.}(2025)]%
        {he2025cameractrlenablingcameracontrol}
\bibfield{author}{\bibinfo{person}{Hao He}, \bibinfo{person}{Yinghao Xu},
  \bibinfo{person}{Yuwei Guo}, \bibinfo{person}{Gordon Wetzstein},
  \bibinfo{person}{Bo Dai}, \bibinfo{person}{Hongsheng Li}, {and}
  \bibinfo{person}{Ceyuan Yang}.} \bibinfo{year}{2025}\natexlab{}.
\newblock \bibinfo{title}{CameraCtrl: Enabling Camera Control for Text-to-Video
  Generation}.
\newblock
\showeprint[arxiv]{2404.02101}~[cs.CV]
\urldef\tempurl%
\url{https://arxiv.org/abs/2404.02101}
\showURL{%
\tempurl}


\bibitem[He et~al\mbox{.}(2023)]%
        {he2023virtual}
\bibfield{author}{\bibinfo{person}{Li-wei He}, \bibinfo{person}{Michael~F
  Cohen}, {and} \bibinfo{person}{David~H Salesin}.}
  \bibinfo{year}{2023}\natexlab{}.
\newblock \showarticletitle{The virtual cinematographer: A paradigm for
  automatic real-time camera control and directing}.
\newblock In \bibinfo{booktitle}{\emph{Seminal Graphics Papers: Pushing the
  Boundaries, Volume 2}}. \bibinfo{pages}{707--714}.
\newblock


\bibitem[He et~al\mbox{.}(2024)]%
        {he2024interactive}
\bibfield{author}{\bibinfo{person}{Rui He}, \bibinfo{person}{Huaxin Wei}, {and}
  \bibinfo{person}{Ying Cao}.} \bibinfo{year}{2024}\natexlab{}.
\newblock \showarticletitle{An Interactive System for Supporting Creative
  Exploration of Cinematic Composition Designs}. In
  \bibinfo{booktitle}{\emph{Proceedings of the 37th Annual ACM Symposium on
  User Interface Software and Technology}}. \bibinfo{pages}{1--15}.
\newblock


\bibitem[Ho et~al\mbox{.}(2022)]%
        {ho2022video}
\bibfield{author}{\bibinfo{person}{Jonathan Ho}, \bibinfo{person}{Tim
  Salimans}, \bibinfo{person}{Alexey Gritsenko}, \bibinfo{person}{William
  Chan}, \bibinfo{person}{Mohammad Norouzi}, {and} \bibinfo{person}{David~J
  Fleet}.} \bibinfo{year}{2022}\natexlab{}.
\newblock \showarticletitle{Video diffusion models}.
\newblock \bibinfo{journal}{\emph{Advances in neural information processing
  systems}}  \bibinfo{volume}{35} (\bibinfo{year}{2022}),
  \bibinfo{pages}{8633--8646}.
\newblock


\bibitem[Hu(2024)]%
        {hu2024animate}
\bibfield{author}{\bibinfo{person}{Li Hu}.} \bibinfo{year}{2024}\natexlab{}.
\newblock \showarticletitle{Animate anyone: Consistent and controllable
  image-to-video synthesis for character animation}. In
  \bibinfo{booktitle}{\emph{Proceedings of the IEEE/CVF Conference on Computer
  Vision and Pattern Recognition}}. \bibinfo{pages}{8153--8163}.
\newblock


\bibitem[Huang et~al\mbox{.}(2025)]%
        {huang2025filmaster}
\bibfield{author}{\bibinfo{person}{Kaiyi Huang}, \bibinfo{person}{Yukun Huang},
  \bibinfo{person}{Xintao Wang}, \bibinfo{person}{Zinan Lin},
  \bibinfo{person}{Xuefei Ning}, \bibinfo{person}{Pengfei Wan},
  \bibinfo{person}{Di Zhang}, \bibinfo{person}{Yu Wang}, {and}
  \bibinfo{person}{Xihui Liu}.} \bibinfo{year}{2025}\natexlab{}.
\newblock \showarticletitle{Filmaster: Bridging cinematic principles and
  generative ai for automated film generation}.
\newblock \bibinfo{journal}{\emph{arXiv preprint arXiv:2506.18899}}
  (\bibinfo{year}{2025}).
\newblock


\bibitem[Huang et~al\mbox{.}(2020)]%
        {huang2020movienet}
\bibfield{author}{\bibinfo{person}{Qingqiu Huang}, \bibinfo{person}{Yu Xiong},
  \bibinfo{person}{Anyi Rao}, \bibinfo{person}{Jiaze Wang}, {and}
  \bibinfo{person}{Dahua Lin}.} \bibinfo{year}{2020}\natexlab{}.
\newblock \showarticletitle{MovieNet: A Holistic Dataset for Movie
  Understanding}. In \bibinfo{booktitle}{\emph{The European Conference on
  Computer Vision (ECCV)}}.
\newblock


\bibitem[Huang et~al\mbox{.}(2022)]%
        {huang2022real}
\bibfield{author}{\bibinfo{person}{Zhewei Huang}, \bibinfo{person}{Tianyuan
  Zhang}, \bibinfo{person}{Wen Heng}, \bibinfo{person}{Boxin Shi}, {and}
  \bibinfo{person}{Shuchang Zhou}.} \bibinfo{year}{2022}\natexlab{}.
\newblock \showarticletitle{Real-time intermediate flow estimation for video
  frame interpolation}. In \bibinfo{booktitle}{\emph{European conference on
  computer vision}}. Springer, \bibinfo{pages}{624--642}.
\newblock


\bibitem[Jhala and Young(2005)]%
        {jhala2005discourse}
\bibfield{author}{\bibinfo{person}{Arnav Jhala} {and}
  \bibinfo{person}{Robert~Michael Young}.} \bibinfo{year}{2005}\natexlab{}.
\newblock \showarticletitle{A discourse planning approach to cinematic camera
  control for narratives in virtual environments}. In
  \bibinfo{booktitle}{\emph{AAAI}}, Vol.~\bibinfo{volume}{5}.
  \bibinfo{pages}{307--312}.
\newblock


\bibitem[Jorgensen et~al\mbox{.}(2023)]%
        {jorgensen2023screenplay}
\bibfield{author}{\bibinfo{person}{Kyle Jorgensen}, \bibinfo{person}{Haohong
  Wang}, {and} \bibinfo{person}{Mea Wang}.} \bibinfo{year}{2023}\natexlab{}.
\newblock \showarticletitle{From screenplay to screen: A natural language
  processing approach to animated film making}. In
  \bibinfo{booktitle}{\emph{2023 International Conference on Computing,
  Networking and Communications (ICNC)}}. IEEE, \bibinfo{pages}{484--490}.
\newblock


\bibitem[Kato et~al\mbox{.}(2024)]%
        {kato2024griffith}
\bibfield{author}{\bibinfo{person}{Jun Kato}, \bibinfo{person}{Kenta Hara},
  {and} \bibinfo{person}{Nao Hirasawa}.} \bibinfo{year}{2024}\natexlab{}.
\newblock \showarticletitle{Griffith: A Storyboarding Tool Designed with
  Japanese Animation Professionals}. In \bibinfo{booktitle}{\emph{Proceedings
  of the 2024 CHI Conference on Human Factors in Computing Systems}}.
  \bibinfo{pages}{1--14}.
\newblock


\bibitem[Katz(1991)]%
        {katz1991film}
\bibfield{author}{\bibinfo{person}{Steven~Douglas Katz}.}
  \bibinfo{year}{1991}\natexlab{}.
\newblock \bibinfo{booktitle}{\emph{Film directing shot by shot: visualizing
  from concept to screen}}.
\newblock \bibinfo{publisher}{Gulf Professional Publishing}.
\newblock


\bibitem[Kim et~al\mbox{.}(2021)]%
        {kim2021asap}
\bibfield{author}{\bibinfo{person}{Hanseob Kim}, \bibinfo{person}{Ghazanfar
  Ali}, {and} \bibinfo{person}{Jae-In Hwang}.} \bibinfo{year}{2021}\natexlab{}.
\newblock \showarticletitle{ASAP: Auto-generating Storyboard And Previz with
  Virtual Humans.}. In \bibinfo{booktitle}{\emph{ISMAR Adjunct}}.
  \bibinfo{pages}{316--320}.
\newblock


\bibitem[Kuleshov and Kuleshov(1974)]%
        {kuleshov1974kuleshov}
\bibfield{author}{\bibinfo{person}{Lev~Vladimirovich Kuleshov} {and}
  \bibinfo{person}{Lev Kuleshov}.} \bibinfo{year}{1974}\natexlab{}.
\newblock \bibinfo{booktitle}{\emph{Kuleshov on film: Writings}}.
\newblock \bibinfo{publisher}{Univ of California Press}.
\newblock


\bibitem[Li et~al\mbox{.}(2024)]%
        {li2024anim}
\bibfield{author}{\bibinfo{person}{Yunxin Li}, \bibinfo{person}{Haoyuan Shi},
  \bibinfo{person}{Baotian Hu}, \bibinfo{person}{Longyue Wang},
  \bibinfo{person}{Jiashun Zhu}, \bibinfo{person}{Jinyi Xu},
  \bibinfo{person}{Zhen Zhao}, {and} \bibinfo{person}{Min Zhang}.}
  \bibinfo{year}{2024}\natexlab{}.
\newblock \showarticletitle{Anim-director: A large multimodal model powered
  agent for controllable animation video generation}. In
  \bibinfo{booktitle}{\emph{SIGGRAPH Asia 2024 Conference Papers}}.
  \bibinfo{pages}{1--11}.
\newblock


\bibitem[Liu et~al\mbox{.}(2026)]%
        {liu2026text}
\bibfield{author}{\bibinfo{person}{Xingyu~Bruce Liu}, \bibinfo{person}{Mira
  Dontcheva}, {and} \bibinfo{person}{Dingzeyu Li}.}
  \bibinfo{year}{2026}\natexlab{}.
\newblock \showarticletitle{A Text-Native Interface for Generative Video
  Authoring}.
\newblock \bibinfo{journal}{\emph{arXiv preprint arXiv:2603.09072}}
  (\bibinfo{year}{2026}).
\newblock


\bibitem[Liveley(2019)]%
        {liveley2019narratology}
\bibfield{author}{\bibinfo{person}{Genevieve Liveley}.}
  \bibinfo{year}{2019}\natexlab{}.
\newblock \bibinfo{booktitle}{\emph{Narratology}}.
\newblock \bibinfo{publisher}{Oxford University Press}.
\newblock


\bibitem[Masson et~al\mbox{.}(2025)]%
        {masson2025visual}
\bibfield{author}{\bibinfo{person}{Damien Masson}, \bibinfo{person}{Zixin
  Zhao}, {and} \bibinfo{person}{Fanny Chevalier}.}
  \bibinfo{year}{2025}\natexlab{}.
\newblock \showarticletitle{Visual Story-Writing: Writing by Manipulating
  Visual Representations of Stories}. In \bibinfo{booktitle}{\emph{Proceedings
  of the 38th Annual ACM Symposium on User Interface Software and Technology}}.
  \bibinfo{pages}{1--15}.
\newblock


\bibitem[McKee(1997)]%
        {mckee1997story}
\bibfield{author}{\bibinfo{person}{Robert McKee}.}
  \bibinfo{year}{1997}\natexlab{}.
\newblock \bibinfo{booktitle}{\emph{Story: style, structure, substance, and the
  principles of screenwriting}}.
\newblock \bibinfo{publisher}{Harper Collins}.
\newblock


\bibitem[Rao et~al\mbox{.}(2024)]%
        {rao2024scriptviz}
\bibfield{author}{\bibinfo{person}{Anyi Rao}, \bibinfo{person}{Jean-Pe{\"\i}c
  Chou}, {and} \bibinfo{person}{Maneesh Agrawala}.}
  \bibinfo{year}{2024}\natexlab{}.
\newblock \showarticletitle{Scriptviz: A visualization tool to aid
  scriptwriting based on a large movie database}. In
  \bibinfo{booktitle}{\emph{Proceedings of the 37th Annual ACM Symposium on
  User Interface Software and Technology}}. \bibinfo{pages}{1--13}.
\newblock


\bibitem[Savardi et~al\mbox{.}(2018)]%
        {savardi2018shot}
\bibfield{author}{\bibinfo{person}{Mattia Savardi}, \bibinfo{person}{Alberto
  Signoroni}, \bibinfo{person}{Pierangelo Migliorati}, {and}
  \bibinfo{person}{Sergio Benini}.} \bibinfo{year}{2018}\natexlab{}.
\newblock \showarticletitle{Shot Scale Analysis in Movies by Convolutional
  Neural Networks}. In \bibinfo{booktitle}{\emph{2018 25th IEEE International
  Conference on Image Processing (ICIP)}}. \bibinfo{pages}{2620--2624}.
\newblock
\href{https://doi.org/10.1109/ICIP.2018.8451474}{doi:\nolinkurl{10.1109/ICIP.2018.8451474}}


\bibitem[Seedream et~al\mbox{.}(2025)]%
        {seedream2025seedream}
\bibfield{author}{\bibinfo{person}{Team Seedream}, \bibinfo{person}{Yunpeng
  Chen}, \bibinfo{person}{Yu Gao}, \bibinfo{person}{Lixue Gong},
  \bibinfo{person}{Meng Guo}, \bibinfo{person}{Qiushan Guo},
  \bibinfo{person}{Zhiyao Guo}, \bibinfo{person}{Xiaoxia Hou},
  \bibinfo{person}{Weilin Huang}, \bibinfo{person}{Yixuan Huang},
  {et~al\mbox{.}}} \bibinfo{year}{2025}\natexlab{}.
\newblock \showarticletitle{Seedream 4.0: Toward next-generation multimodal
  image generation}.
\newblock \bibinfo{journal}{\emph{arXiv preprint arXiv:2509.20427}}
  (\bibinfo{year}{2025}).
\newblock


\bibitem[Soucek and Lokoc(2024)]%
        {soucek2024TransNetv2}
\bibfield{author}{\bibinfo{person}{Tom\'{a}s Soucek} {and}
  \bibinfo{person}{Jakub Lokoc}.} \bibinfo{year}{2024}\natexlab{}.
\newblock \showarticletitle{TransNet V2: An Effective Deep Network Architecture
  for Fast Shot Transition Detection}. In \bibinfo{booktitle}{\emph{Proceedings
  of the 32nd ACM International Conference on Multimedia}} (Melbourne VIC,
  Australia) \emph{(\bibinfo{series}{MM '24})}. \bibinfo{publisher}{Association
  for Computing Machinery}, \bibinfo{address}{New York, NY, USA},
  \bibinfo{pages}{11218–11221}.
\newblock
\showISBNx{9798400706868}
\href{https://doi.org/10.1145/3664647.3685517}{doi:\nolinkurl{10.1145/3664647.3685517}}


\bibitem[Suh et~al\mbox{.}(2024)]%
        {suh2024luminate}
\bibfield{author}{\bibinfo{person}{Sangho Suh}, \bibinfo{person}{Meng Chen},
  \bibinfo{person}{Bryan Min}, \bibinfo{person}{Toby Jia-Jun Li}, {and}
  \bibinfo{person}{Haijun Xia}.} \bibinfo{year}{2024}\natexlab{}.
\newblock \showarticletitle{Luminate: Structured generation and exploration of
  design space with large language models for human-ai co-creation}. In
  \bibinfo{booktitle}{\emph{Proceedings of the 2024 CHI Conference on Human
  Factors in Computing Systems}}. \bibinfo{pages}{1--26}.
\newblock


\bibitem[Sweller(1988)]%
        {sweller1988cognitive}
\bibfield{author}{\bibinfo{person}{John Sweller}.}
  \bibinfo{year}{1988}\natexlab{}.
\newblock \showarticletitle{Cognitive load during problem solving: Effects on
  learning}.
\newblock \bibinfo{journal}{\emph{Cognitive science}} \bibinfo{volume}{12},
  \bibinfo{number}{2} (\bibinfo{year}{1988}), \bibinfo{pages}{257--285}.
\newblock


\bibitem[Tian(2025)]%
        {tian2025large}
\bibfield{author}{\bibinfo{person}{Ke Tian}.} \bibinfo{year}{2025}\natexlab{}.
\newblock \showarticletitle{A Large Language Model-Based System for Semantic
  Understanding and Automated Scene Generation in Animation Scripts}. In
  \bibinfo{booktitle}{\emph{Proceedings of the 2nd International Conference on
  Machine Intelligence and Digital Applications}}. \bibinfo{pages}{116--120}.
\newblock


\bibitem[Wan et~al\mbox{.}(2025)]%
        {wan2025wan}
\bibfield{author}{\bibinfo{person}{Team Wan}, \bibinfo{person}{Ang Wang},
  \bibinfo{person}{Baole Ai}, \bibinfo{person}{Bin Wen},
  \bibinfo{person}{Chaojie Mao}, \bibinfo{person}{Chen-Wei Xie},
  \bibinfo{person}{Di Chen}, \bibinfo{person}{Feiwu Yu},
  \bibinfo{person}{Haiming Zhao}, \bibinfo{person}{Jianxiao Yang},
  {et~al\mbox{.}}} \bibinfo{year}{2025}\natexlab{}.
\newblock \showarticletitle{Wan: Open and advanced large-scale video generative
  models}.
\newblock \bibinfo{journal}{\emph{arXiv preprint arXiv:2503.20314}}
  (\bibinfo{year}{2025}).
\newblock


\bibitem[Wei et~al\mbox{.}(2025)]%
        {wei2025cinevision}
\bibfield{author}{\bibinfo{person}{Zheng Wei}, \bibinfo{person}{Hongtao Wu},
  \bibinfo{person}{Lvmin Zhang}, \bibinfo{person}{Xian Xu},
  \bibinfo{person}{Yefeng Zheng}, \bibinfo{person}{Pan Hui},
  \bibinfo{person}{Maneesh Agrawala}, \bibinfo{person}{Huamin Qu}, {and}
  \bibinfo{person}{Anyi Rao}.} \bibinfo{year}{2025}\natexlab{}.
\newblock \showarticletitle{CineVision: An Interactive Pre-visualization
  Storyboard System for Director--Cinematographer Collaboration}. In
  \bibinfo{booktitle}{\emph{Proceedings of the 38th Annual ACM Symposium on
  User Interface Software and Technology}}. \bibinfo{pages}{1--18}.
\newblock


\bibitem[Yeh et~al\mbox{.}(2026)]%
        {yeh2026vidmento}
\bibfield{author}{\bibinfo{person}{Catherine Yeh}, \bibinfo{person}{Anh
  Truong}, \bibinfo{person}{Mira Dontcheva}, {and} \bibinfo{person}{Bryan
  Wang}.} \bibinfo{year}{2026}\natexlab{}.
\newblock \showarticletitle{Vidmento: Creating Video Stories Through
  Context-Aware Expansion With Generative Video}.
\newblock \bibinfo{journal}{\emph{arXiv preprint arXiv:2601.22013}}
  (\bibinfo{year}{2026}).
\newblock


\bibitem[Yu et~al\mbox{.}(2024)]%
        {yu2024barriers}
\bibfield{author}{\bibinfo{person}{Tao Yu}, \bibinfo{person}{Wei Yang},
  \bibinfo{person}{Junping Xu}, {and} \bibinfo{person}{Younghwan Pan}.}
  \bibinfo{year}{2024}\natexlab{}.
\newblock \showarticletitle{Barriers to industry adoption of AI video
  generation tools: A study based on the perspectives of video production
  professionals in China}.
\newblock \bibinfo{journal}{\emph{Applied Sciences}} \bibinfo{volume}{14},
  \bibinfo{number}{13} (\bibinfo{year}{2024}), \bibinfo{pages}{5770}.
\newblock


\bibitem[Zhang et~al\mbox{.}(2025)]%
        {zhang2025bridging}
\bibfield{author}{\bibinfo{person}{Jiaxu Zhang}, \bibinfo{person}{Tianshu Hu},
  \bibinfo{person}{Yuan Zhang}, \bibinfo{person}{Zenan Li},
  \bibinfo{person}{Linjie Luo}, \bibinfo{person}{Guosheng Lin}, {and}
  \bibinfo{person}{Xin Chen}.} \bibinfo{year}{2025}\natexlab{}.
\newblock \showarticletitle{Bridging Your Imagination with Audio-Video
  Generation via a Unified Director}.
\newblock \bibinfo{journal}{\emph{arXiv preprint arXiv:2512.23222}}
  (\bibinfo{year}{2025}).
\newblock


\bibitem[Zhou et~al\mbox{.}(2024a)]%
        {zhou2024visar}
\bibfield{author}{\bibinfo{person}{Juntao Zhou}, \bibinfo{person}{Yijie Li},
  \bibinfo{person}{Yida Wang}, \bibinfo{person}{Dian Ding}, \bibinfo{person}{Yu
  Lu}, \bibinfo{person}{Yi-Chao Chen}, {and} \bibinfo{person}{Guangtao Xue}.}
  \bibinfo{year}{2024}\natexlab{a}.
\newblock \showarticletitle{Visar: Projecting virtual sound spots for acoustic
  augmented reality using air nonlinearity}.
\newblock \bibinfo{journal}{\emph{Proceedings of the ACM on Interactive,
  Mobile, Wearable and Ubiquitous Technologies}} \bibinfo{volume}{8},
  \bibinfo{number}{3} (\bibinfo{year}{2024}), \bibinfo{pages}{1--30}.
\newblock


\bibitem[Zhou et~al\mbox{.}(2024b)]%
        {zhou2024storydiffusion}
\bibfield{author}{\bibinfo{person}{Yupeng Zhou}, \bibinfo{person}{Daquan Zhou},
  \bibinfo{person}{Ming-Ming Cheng}, \bibinfo{person}{Jiashi Feng}, {and}
  \bibinfo{person}{Qibin Hou}.} \bibinfo{year}{2024}\natexlab{b}.
\newblock \showarticletitle{Storydiffusion: Consistent self-attention for
  long-range image and video generation}.
\newblock \bibinfo{journal}{\emph{Advances in Neural Information Processing
  Systems}}  \bibinfo{volume}{37} (\bibinfo{year}{2024}),
  \bibinfo{pages}{110315--110340}.
\newblock


\end{thebibliography}

\appendix

\end{document}